\newcommand {\avrg}[1] {\mbox{$\langle\, #1\rangle$}}

\newcommand {\Mpc}   {\mbox{$ h^{-1}$} Mpc\,}

\documentstyle{l-aa}

\begin{document}

\thesaurus{
       (12.03.3;  
        12.12.1;  
        11.03.1;  
        11.11.1)  
          }

\title{The ESO Nearby Abell Cluster Survey
       \thanks{Based on observations collected at the European Southern
               Observatory (La Silla, Chile)}
      }

\subtitle{I. Description of the Dataset and Definition of Physical Systems
       \thanks{Table 6 is also available in electronic form at the CDS via
               anonymous ftp 130.79.128.5}}

\author{P.~Katgert \inst{1}, A.~Mazure \inst{2},
  J.~Perea \inst{3}, R.~den Hartog \inst{1},
  M.~Moles \inst{3}, O.~Le~F\`evre \inst{4},
  P.~Dubath \inst{5}, P.~Focardi \inst{6},
  G.~Rhee \inst{7}, B.~Jones \inst{8},
  E.~Escalera \inst{9,10}, A.~Biviano \inst{1,11},
  D.~Gerbal \inst{4,11}, \and G.~Giuricin \inst{9,12}
       }

\institute{Sterrewacht Leiden, Netherlands \and
           Laboratoire d'Astronomie Spatiale, Marseille, France \and
           Instituto de Astrof\'{\i}sica de Andaluc\'{\i}a, CSIC, Granada,
Spain \and
           DAEC, Observatoire de Paris, Universit\'e Paris 7, CNRS (UA 173),
France \and
           Observatoire de Gen\`eve, Switzerland \& Lick Observatory, Univ. of
California, Santa Cruz, USA \and
           Dipartimento di Astronomia, Universit\`a di Bologna, Italy \and
           Department of Physics, University of Nevada, Las Vegas, USA \and
           Nordita, Copenhagen, Denmark \and
           Dipartimento di Astronomia, Universit\`a di Trieste, Italy \and
	   Royal Observatory, Edinburgh, UK \and
           Institut d'Astrophysique de Paris, France \and
           SISSA, Trieste, Italy
           }

\offprints{P.~Katgert}
\date{Received date; accepted date}

\maketitle
\markboth{The ESO Nearby Abell Cluster Survey: I. Dataset and Definition of
Physical Systems}{}

\begin{abstract}

We describe the results of the ESO Key-programme on ``Structure and
Dynamics of Rich Galaxy Clusters'' (which we will henceforth refer to
as the ESO Nearby Abell Cluster Survey - or ENACS). We discuss the
sample of clusters for which data were obtained, and the
observational programme of spectroscopy and photometry that we
carried out. The final database contains a total of 5634 galaxies in
the directions of 107 clusters from the catalogue by Abell, Corwin
and Olowin 1989 (ACO hereafter) with richness $R\ge 1$ and mean
redshifts $z \le 0.1$. For 4465 galaxies the redshift is based solely
on absorption lines, for 586 galaxies it is based on both absorption
and emission lines, while for the remaining 583 galaxies the redshift
is based exclusively on one or more emission lines. For 5615 galaxies
an $R_{25}$ magnitude was obtained. We discuss in some detail the
methods of observation and analysis and determine the quality and
reliability of the data from independent, repeated measurements. All
absorption-line redshifts with a low S/N-ratio of the peak in the
correlation function have been judged on plausibility by combined
visual inspection of spectrum and correlation function. This has led
to an empirically determined overall reliabilty of the 5634 accepted
redshifts of 0.98.

We discuss various methods for defining the compact, physically
relevant systems in the 107 `pencil-beam' surveys. We have chosen to
apply a fixed velocity gap to separate galaxies that do not form part
of the same system. From the summed distribution of the velocity
differences between galaxies adjacent in redshift we conclude that,
for the average system in our survey, and for our sampling of the
velocity distributions, a velocity gap of 1000 km/s is the optimum
one for defining the systems. With this gap size, systems are not
broken up into sub-systems, and field galaxies are hardly linked to
the systems. We present the mean redshifts of the 220 systems (with
at least 4 members) identified in the 107 pencil beam surveys, using
a fixed gap of 1000 km/s.

On average, about 75\% of the 5634 galaxies are in the largest system
found in the direction of the rich Abell cluster candidate. This
shows that, within $\approx 1$ \Mpc of the cluster centre and down to
$R \approx 17$, field contamination is not negligible for clusters
with $z \la 0.1$. However, about half of the 25\% of galaxies outside
the largest system belong to a secondary system along the same line
of sight.

At the same time, field contamination has produced only a small
number of spurious rich clusters within the sample of $R\ge 1$ , $z
\le 0.1$ clusters. For about 90 \% of the nearby rich Abell cluster
candidates studied here we find a redshift system that either
contains more than half of the total number of redshifts, or that has
at least two times as many redshifts as the next largest system. Only
in about 10\% of the cases does an $R\ge 1$ , $z \le 0.1$ entry in
the ACO catalogue appear to be the result of a superposition of two
almost equally rich (but relatively poorer) systems. Almost all of
the rich and relatively nearby ACO cluster candidates that we studied
thus appear to be real rich clusters that represent physical systems.

\end{abstract}


\section{Introduction}

Rich clusters of galaxies have long been recognized as objects that
can provide information about important aspects of the physics of
large-scale structure formation. Their spatial distribution, their
motions with respect to the Hubble flow, and the distribution of
their global properties, such as total mass and shape, all hold clues
to details of the formation process. Among the latter are the form
and amplitude of the spectrum of initial fluctuations on cluster
scales, and the average density of the expanding background Universe
in which the clusters form (e.g.\ White 1992, and references therein).

In the central parts of rich clusters, the memory of the initial
conditions on sub-cluster scales has most probably been erased
completely. On the other hand, the relaxation times outside the cores
are definitely longer than a Hubble-time so that (dynamical)
structure in the outer parts may contain clues about initial
conditions on sub-cluster scales, and about dynamical processes that
are important during the collapse of the cluster (e.g.\ West, Dekel \&
Oemler 1987). In addition, the central regions of rich clusters are
ideal `laboratories' for the study of the dynamical effects that
occur in environments where the galaxy density is very much higher
than average. Finally, the present-day distribution of the masses of
the cluster galaxies reflects the initial mass function on galaxy
scales, modified by the growth and destruction processes that result
from encounters between galaxies (e.g.\ Sarazin 1986).

For several aspects of the study of the properties of rich clusters
it suffices to observe a well-chosen set of clusters, without the
need to rigorously define a complete sample. However, for many
purposes it is essential that the analysis is based on a cluster
sample that is statistically complete. Obvious cases include: the
determination of the cluster mass function, and the study of the
distribution of cluster shapes. In these instances, a useful
comparison with model predictions is only possible if the dataset
that is used to constrain the models is complete in a well-defined
manner, or if its incompleteness can be described with sufficient
accuracy.

Complete samples of clusters can at present be defined in two ways.
Either one uses optical galaxy catalogues, or high-latitude surveys
of X-ray sources to define candidate clusters. In both cases
spectroscopic follow-up is required to confirm the reality of the
candidate clusters and to find their distances. In the near future it
will become possible to use wide-area redshift surveys of galaxies
out to considerable distances to define cluster samples in a
direct and controlled manner.

Both of the methods presently available to define cluster samples
have their disadvantages. Optical cluster catalogues, whether based
on visual inspection of survey plates (Abell 1958, and Abell, Corwin
\& Olowin 1989, hereafter ACO) or on galaxy catalogues generated
with automatic scanning machines (Lumsden et al.\ 1992, Dalton et
al.\ 1992) suffer from superposition effects; i.e.\ the spatial
compactness of the peaks in the projected galaxy distribution is not
guaranteed, and has to be confirmed. Moreover, the only parameter
with respect to which such a cluster sample can be defined to be
complete is apparent richness: the number of member galaxies in a
cone with a fixed cross-section at the distance of the cluster in a
specified range of apparent magnitude. The relation between richness
and a physical property such as total mass is not obvious, so
completeness in terms of total mass is much harder to achieve.

The superposition problem hardly exists for cluster samples based
on X-ray surveys, because in X-rays the contrast of a cluster with
respect to the field is much higher than it is in the galaxy
distribution. However, samples of X-ray clusters can be
made complete only with respect to X-ray flux. The extraction of a
volume-limited sample (as required for several types of argument)
requires spectroscopy of member galaxies, which then also yields
X-ray luminosities (e.g.\ Briel and Henry 1993, and Pierre et al.\
1994). Even X-ray based cluster samples may not be complete with
respect to a physically relevant parameter like total mass, if X-ray
luminosity and total mass are not very strongly correlated.

There now exist several extensive redshift surveys of galaxies in
clusters in the literature (e.g.\ Colless \& Hewett 1987, Dressler \&
Shectman 1988, Teague, Carter \& Gray 1990,  Zabludoff, Huchra \&
Geller 1990, ZHG hereafter, Malumuth et al.\ 1992, Guzzo et al.\
1992, Dalton et al.\ 1994). However, these data do not yet allow the
construction of a large complete, volume-limited sample of rich
clusters with good data on e.g.\ velocity dispersions, internal
structure etc. (even if some of the surveys are complete in one way
or another). In this paper we describe an observational project,
with the status of an ESO Key-programme, which is aimed at providing
good and extensive redshift data for a complete sample of at least
100 rich clusters out to a redshift of about 0.1. In combination with
literature data this should yield accurate mean redshifts for well
over 100 clusters and meaningful estimates of global velocity
dispersion for a large fraction of those. For a subset of between 20
and 30 of the richest clusters, the aim is to obtain at least about
100 redshifts for a detailed analysis of the kinematics of the
cluster galaxies, to allow a discussion of the cluster dynamics.

The impact of our programme in this area can be illustrated as
follows. A recent compilation by Biviano et al.\ (1992) of
available cluster redshift data in the literature contains 6470
redshifts; hence: the ENACS almost doubles the available number of
redshifts in the direction of rich clusters. More importantly: it is
the largest homogeneous dataset of redshifts of galaxies in clusters.
This is clear from our discussion of the distribution of velocity
dispersions for a complete, volume-limited cluster sample (Mazure et
al.\ 1995). In that case more than 80\% of the redshifts were
contributed by our redshift survey.

In Section 2 we describe the cluster sample that we have studied, as
well as the definition of the galaxy samples in the direction of
these clusters, on which we carried out multi-object spectroscopy. In
Section 3 we briefly describe the spectroscopic observations, some
relevant details of the spectroscopic reduction, and the methods by
which we obtained absorption- and emission-line redshifts. In Section
4 we discuss the quality and the reliability of the redshift
estimates, internally from multiple measurements, and externally from
comparison with literature data. In Section 5 we describe how we
calibrated our photographic photometry with CCD-imaging, and we discuss
the quality of the calibration, and the completeness of the redshift
surveys. In Section 6 we discuss the results of our spectroscopy,
from the point of view of the definition of physically relevant
redshift systems in our pencil-beam surveys which are centered on
target clusters from the ACO catalogue. In Section 7 we discuss for
our cluster sample the effects of field contamination and
superposition. In Section 8 we describe the spatial distribution of
the clusters, and of the galaxies in the clusters for which the
positional selection function is not straightforward. Finally, in
Section 9 we summarize the most important conclusions.

\section{The Cluster Sample and the Definition of the Galaxy Samples}

\subsection{The Sample of Clusters observed in the ENACS}

We designed the ENACS so that it would establish, upon completion and
in combination with literature data, a redshift database for an
essentially complete sample of Abell clusters with richness $R \ge 1$
out to a redshift $z$ of about 0.1, in the solid angle of 2.55 sr
around the South Galactic Pole, defined by $b \le - 30\degr$ and
$-70\degr \le \delta \le 0\degr$. We included all $R \ge 1$ clusters
which, at the start of the project in 1989, either had a
spectroscopic redshift $z \le 0.1$ or had a high probability of
having $z \le 0.1$, in view of their value of $m_{10}$. As the
$m_{10} - z$ relation indicates that most of the clusters with $z \le
0.1$ have $m_{10} \le 16.9$, we included all clusters with $m_{10}
\le 16.9$. The width of the $m_{10} - z$ relation implies that one
will not have included {\em all} clusters with $z \le 0.1$ within
this $m_{10}$ limit. To become truly complete out to $z = 0.1$ would
have required a (much) fainter limit in $m_{10}$. Then, however, the
scope of the project would have increased unacceptably, while the
problem of the lack of a precise completeness limit in redshift would
not have disappeared but only have shifted to a higher redshift.

In the course of the project we took into account all information
that became available to update the list of clusters still to be
observed; i.e.\ we tried to minimize duplication with work being done
elsewhere, and to optimize the yield of our project. We also used the
interim results from our observations to improve the definition of
the subset of clusters for which we tried to obtain of order 100
redshifts (in our jargon a `structure cluster'). The `promotion' of a
cluster into the structure-cluster category only took place if the
system was sufficiently rich and relatively compact in redshift
space, without significant secondary systems at other redshifts.

The full history of the evolution of the definition of the sample is
not of interest here. In Section 6 the outcome of the process is
summarized in the form of a list of the Abell clusters that we have
observed, and a global description of the redshift distribution that
we measured in the direction of each cluster.

In addition to the clusters in the region defined above, we also
observed a few of the clusters in the so-called Shapley concentration
(e.g.\ Bardelli et al.\ 1994), which is in the Northern galactic
hemisphere, around $\alpha = 13^{\rm h}$. In order to make best use
of telescope time we added a few `filler' clusters around $\alpha =
12^{\rm h}$ with $\delta > 0 \degr$.

\subsection{Definition of the Galaxy Samples for Spectroscopy}

The Southern galaxy catalogues that have been produced in the last
few years by groups in the UK, using automatic plate-scanning
machines (Lumsden et al.\ 1992, and Dalton et al.\ 1992) were not yet
available at the beginning of our project. For that reason we had to
produce `special-purpose' galaxy catalogues around the target
clusters ourselves. We used the Leiden Observatory {\sl Astroscan}
plate-measuring machine (de Vries 1987, Swaans 1981, Van Haarlem et
al.\ 1991) to produce such catalogues. Where possible, we used film
copies of the SERC blue survey (IIIa-J emulsion). If that was not
possible, the Leiden Observatory glass copies of the first Palomar
Sky Survey (103a-E emulsion) were used. Typically, areas of between
about 1 and 4 square degrees were scanned at 10 micron ($\approx 0.6
\arcsec$) resolution.
The threshold for object detection was set at a level above sky of
between 5 and 7 times the sky noise. For the IIIa-J survey plates
this corresponds roughly to a surface brightness cut-off at the 22
mag/arcsec$^2$ isophote, while for the 103a-E emulsion the cut-off
corresponds to about 20.5 mag/arcsec$^2$.

The catalogues of objects detected above the threshold contain
between several and many thousands of entries per cluster. For each
of the objects a few basic parameters were determined on-line. The
most important ones are the centre-of-gravity position, a photometric
parameter $P_{\rm phot}$ (which measures the brightness of the
object), and the second moment (size) of the image. The photometric
parameter corresponds to `the amount of silver' in the thresholded
part of the image and its logarithm shows an almost linear relation
with magnitude (see Section 5).

Of course, many objects in the catalogues are not galaxies. In order
to eliminate stellar objects as much as possible, we used a two-step
scheme. First, we applied an automatic star-galaxy discriminator
which uses the second moment of the object in combination with its
photometric parameter. Over most of the range in $P_{\rm phot}$ there
is a well-defined, very narrow relation between the size and
brightness of stellar objects. We fitted a Gaussian to the stellar
size distribution in several tens of narrow intervals of brightness.
This allowed us to define non-stellar objects as those deviating by
more than three standard deviations from the median stellar size for
the brightness of the object. Since the stellar locus is very narrow,
the fraction of single stars among the non-stellar objects is very
small.

However, a non-negligible fraction of the (non-stellar) galaxy
candidates are actually double stars. They appear non-stellar because
the two images of the stars merge above the detection threshold. On
the other hand, bright compact saturated galaxies can be
misclassified as stars. In order to ensure that the galaxy catalogues
that we used as the basis of our spectroscopy are not contaminated by
stars at the level of more than a few percent and do contain all the
bright galaxies, we have visually inspected all objects classified as
non-stellar, and all bright objects independent of the
classification. This is clearly a method that is less sophisticated
than the methods employed for the all-sky catalogues (Heydon-Dumbleton et
al.\ 1989, Maddox et al.\ 1990). However, visual pattern recognition
is quite a powerful tool, as confirmed by the results of our
spectroscopy. Contamination of our galaxy catalogues by stars has
been limited to an acceptable level of only a few percent.

It must be stressed that, when we selected galaxies for spectroscopic
observation, the photographic photometry had not yet been calibrated.
The parameter $P_{\rm phot}$ is, however, a sufficiently accurate and
monotonically varying function of magnitude. Therefore, we have
produced galaxy catalogues with well-defined magnitude limits, by
applying a cut-off in $P_{\rm phot}$. This means that we have
attempted spectroscopy for all galaxies within circular areas of
about 0.5$^\circ$ diameter down to well-defined magnitude limits,
which are between 16.5$^m$ and 17.5$^m$ in the R-band. Although the
limit varies between clusters, within a cluster the uniformity of the
limit over the area of the cluster is quite good, even for clusters
for which several fields were observed.

The S/N-ratio in a galaxy spectrum depends primarily on the surface
brightness of that part of the galaxy that illuminates the fibre
entrance. As the relation between isophotal magnitude and (central)
surface brightness of galaxies has an appreciable width, the
spectroscopy cannot be (and is not) complete to a limiting isophotal
magnitude. Our multi-object spectroscopy is, in principle, complete
with respect to central surface brightness within the restriction of
the limit in isophotal magnitude that we imposed. The limit in
central surface brightness is, however, not very sharp as a result of
differences in fibre transmission and in absorption-line strengths.

\section{The Spectroscopic Observations and the Determination of
         Redshifts}

\begin{table*}
\caption[]{Details of the instrumental set-up for the Optopus
spectroscopy with the 3.6-m telescope.}
\begin{flushleft}
\begin{tabular}{llllllll}
\hline\hline
\noalign{\smallskip}
\multicolumn{2}{l}{Observing period} &
\multicolumn{2}{l}{CCD} &
\multicolumn{3}{l}{Spectral} &
\# fibres \\
& & format & pixel & disp.\ & sampl.\ & range & \\
\noalign{\smallskip}
                & year &        & ($\mu$m) & (\AA/mm) & (\AA/pxl) & (\AA) & \\
\noalign{\smallskip}
\hline
\noalign{\smallskip}
03/09 $-$ 08/09 & 1989 & 640$\times$1024 & 15 & 133 & 2.0 & 3855 $-$ 5880 & 31
\\
                &      & 640$\times$1024 & 15 & 172 & 2.6 & 3870 $-$ 6510 & 31
\\
31/03 $-$ 02/04 & 1990 & 640$\times$1024 & 15 & 130 & 1.9 & 3925 $-$ 5860 & 50
\\
14/09 $-$ 17/09 & 1990 & 640$\times$1024 & 15 & 130 & 1.9 & 3930 $-$ 5865 & 50
\\
12/10 $-$ 14/10 & 1990 & 512$\times$512  & 27 & 130 & 3.5 & 3925 $-$ 5725 & 50
\\
02/10 $-$ 08/10 & 1991 & 512$\times$512  & 27 & 130 & 3.5 & 3940 $-$ 5740 & 50
\\
25/09 $-$ 28/09 & 1992 & 512$\times$512  & 27 & 130 & 3.5 & 3870 $-$ 5670 & 50
\\
19/10 $-$ 23/10 & 1992 & 512$\times$512  & 27 & 130 & 3.5 & 3850 $-$ 5650 & 50
\\
14/09 $-$ 16/09 & 1993 & 512$\times$512  & 27 & 130 & 3.5 & 3980 $-$ 5780 & 50
\\
19/10 $-$ 23/10 & 1993 & 512$\times$512  & 27 & 130 & 3.5 & 3980 $-$ 5780 & 50
\\
\noalign{\smallskip}
\hline\hline
\end{tabular}
\end{flushleft}
\end{table*}

\subsection{The Optopus Observations}

All spectroscopic observations were carried out with the Optopus
multi-fibre spectroscopic facility at the ESO 3.6-m telescope at La
Silla (Lund 1986, and Avila et al.\ 1989). This facility employs an
aperture plate at the Cassegrain focus (scale 7.12$\arcsec$/mm), with
a diameter of 33$\arcmin$ (or 274 mm). The system uses fibres with
320 micron (or 2.3$\arcsec$) diameter. At one end of the fibre bundle
each fibre has its own precision `connector' which fits tightly into
holes drilled in the aperture plate at the positions of the objects
to be studied.  The other end of the fibre bundle is formed into a
linear array of fibres fitted into a connector which positions the
fibre ends, which thus replace and define the slit of the Boller \&
Chivens spectrograph.

In Table~1 we list the relevant instrumental details for the various
observing runs. From this Table, some of the modifications that were
made to the system during the period over which the observations were
made (from September 1989 to October 1993), and which improved its
efficiency, can be deduced.

First, the number of fibres increased from 31 to 50 at the beginning
of 1990. At the same time a faster camera was installed in the
spectrograph. Second, in the fall of 1990 a large fraction of the
overhead associated with inserting the 50 fibres in the aperture
plate at the telescope was eliminated by the introduction of a
second, exchangeable fibre bundle (not visible from Table~1, but very
important). This allowed the preparation of an aperture plate during
the exposure of the preceding plate. Third, there has been a marked
improvement in the properties of the CCD-detectors; our project has
benefitted particularly from the decrease of the read-out noise.

The frames were generally exposed for between 60 and 100 minutes,
except in the first observing period, when exposure times were
generally limited to between 35 and 60 minutes which, with the lower
number of fibres (31 instead of 50), causes the observations of
September 1989 to be significantly less efficient than the other
ones.

In general, atmospheric conditions were good during the exposures. A
fair fraction of the exposures was affected by moon-light. In those
cases one cannot use an automatic analysis to find the peak in the
cross-correlation function (see below), as there frequently is a
dominant peak at zero-redshift. Several exposures were carried out
during mediocre, and even cloudy, conditions, which affected the
S/N-ratios in the galaxy spectra quite badly. This applies in
particular to exposures of the clusters A2502, A2933, A3144, A3301
and A3781. As a result, the absence of clear, coherent redshift
systems in the data of A2502 and A3144 (see Section 6) does not prove
conclusively that the target Abell cluster does not exist. We should
mention in this context that for none of the rather short exposures
from September 1989 such ambiguity exists, because in all clusters
that were observed during that run coherent systems were found.

\begin{figure*}
\vbox{}
\caption[]{
An example of a CCD frame obtained with the ESO Optopus multi-object
spectrograph at the ESO 3.6m telescope. Wavelength increases from
left to right. The main absorption and emission features are
indicated. At the far right is the sky line at 5577 \AA.}
\end{figure*}

\subsection{The Accuracy of the Positioning of the Fibres}

In multi-object spectroscopy with aperture plates, the quality of the
positioning of the fibres must be ensured in the preparation of the
aperture plates, as no changes can be made at the telescope.  Because
corrections for differential refraction are applied to the object
positions, the plate is optimally suited for a particular hour angle,
and thus only usable within a window of a few hours around it. The
correct positioning of individual fibres is important; not only does
it guarantee maximum output, but it also minimizes biases due to
possible a-symmetric contributions from galaxy rotation to the
systemic velocities.

The positioning of a plate with respect to the sky employs guide
stars, which allow one to find the correct telescope pointing
position, as well as to align the plate by rotating it. In the
original Optopus system, the pointing and rotation were
simultaneously fixed by two guide stars. An important improvement
resulted from decoupling the two by using a single star (close to the
centre of the plate) for pointing, and a set of guide stars near the
edge of the plate to control rotation. Only guide stars with low
proper motions (i.e.\ $\mu < 0.025\arcsec$/yr) were used. The
pointing errors are estimated to be of the order of at most a few
times 0.1$\arcsec$, while misalignment through rotation does not add
more than a few times 0.1$\arcsec$ at the edge of the plate.

Consistency between the positional systems of guide stars and
galaxies is ensured because all positions are derived from the same
machine scans. For the mapping of the geometry of the survey Schmidt
plates onto the geometry of the 3.6-m Cassegrain focal plane we used
the positions of several tens of standard stars. This mapping takes
into account the effects of deformation and differential refraction
during the exposures of the Schmidt telescope survey plates. From the
quality of the standard star fits we estimate that relative positions
of the fibres are accurate to within again a few times 0.1$\arcsec$.
Finally, the accuracy of the computer-controlled drilling machine, in
combination with the scale of about 7$\arcsec$/mm at Cassegrain
focus, ensures that the mechanical production of the plates does not
corrupt the quality of the positioning system.

\subsection{The Calibration and Reduction of the Optopus Frames}

An exposure with the Optopus system yields a CCD-frame on which about
50 (or 31) parallel, simultaneously recorded spectra are present.
As can be seen from Fig.~1 the extraction of the individual spectra
does not pose any problems, because adjacent spectra are separated by
about 3 non-exposed pixels. The calibration and reduction of the Optopus
CCD-frames is in many ways standard. In particular, in the wavelength
calibration, the dispersion relation was determined for one of the
fibres from the arc-lamp spectra taken before and after each exposure.
Subsequently, for the other fibres zero-point shifts were determined
using this dispersion relation. The quality of the wavelength
calibration was estimated, from the formal quality of the fit (based
mostly on between five and ten arc lines) and from the observed
wavelengths of (mostly two) sky lines, to be between 0.1 and 0.2 \AA.

{}From the extracted wavelength-calibrated galaxy spectra, cosmic-ray
events and emission- and sky-lines were interactively removed. As
shown by Lissandrini et al. (1994) it is possible to achieve
fairly accurate sky subtraction in multi-fibre spectroscopy by using
the strength of one or more sky-lines as a reference. We have
monitored the sky brightness by exposing one or two fibres to blank
sky in each Optopus exposure. From this we found that the sky
brightness is significantly less (over the wavelength range of our
observations) than the average brightness, within the fibre aperture,
of most of our galaxies. Therefore, we decided not to attempt to
subtract the contribution from sky.

Each spectrum was resampled on a regular grid in a logarithmic
wavelength scale, as were several template spectra of bright nearby
galaxies, as well as of several bright galaxies in the clusters that
we investigated, and of stars. Redshifts were determined from
cross-correlating the observed galaxy spectrum with a template
spectrum, using the methods described by Tonry \& Davis (1979). In
particular, we also subtracted (before cross-correlation) a
polynomial fit to the continuum and apodized the result with a cosine
bell to remove the discontinuity between beginning and end of the
spectrum. As is well-known, the correlation strength depends very
much on the extent to which the object and template spectra agree in
detail. After various tests we decided to use a single template to
measure redshift values, namely a spectrum of the nucleus of M31
obtained by Keel at KPNO. We chose this spectrum because it was found
to provide the highest {\sl average} correlation strength. Although
our use of a single template does not provide maximum correlation
strength for each and every galaxy, it does guarantee consistency of
redshift estimates between clusters. Because we have visually
inspected all spectra and correlation functions (which sometimes led
to accepting redshift estimates with low correlation strength) we are
confident to have missed only very few, if any, redshift estimates as
a result of template mismatch.

The positions of the peaks in the correlation function were found
from fitting a parabola to the 5 points around the peak. For each
peak, we determined the S/N-ratio (the R-parameter defined by Tonry
and Davis), while the uncertainty in the redshift estimate was found
from the noise in the correlation function and the curvature of the
peak. As already mentioned, each spectrum was visually inspected,
together with its correlation function. The reality of each redshift
was then judged from the positions of the major absorption lines,
indicated on the spectrum for the redshift corresponding to a
particular peak in the cross-correlation spectrum.

In Fig.~2 we show the distribution of the 5070 absorption-line
redshift estimates that were accepted (of which a few were later
rejected, from a comparison with emission-line redshifts, see
Section 4.3) with respect to the S/N-ratio of the correlation peak,
and redshift. The non-uniform redshift distribution is the result of
the superposition of more than 100 clusters. The detailed analysis of
the galaxy redshift distributions in our survey will be the subject
of a separate paper.

\begin{figure}[t]
\vbox{}
\caption[]{The S/N-ratio of the peak in the cross-correlation function
          against redshift for the 5070 galaxies
          for which an absorption-line redshift was determined.}
\end{figure}

In a completely independent effort, all Optopus CCD-frames were
inspected for the presence of emission lines. The advantage of
inspecting the frames rather than the extracted spectra is that
emission-lines have far `softer' and `rounder' images than cosmic-ray
events, and can thus be distinguished much better from the latter in
the frames than in the one-dimensional spectra. At this point, no
information on the absorption-line redshift was used. I.e.\ the
inspection of the spectra was not limited to those wavelength
intervals where emission lines could be expected, or to those spectra
for which an absorption-line redshift had been obtained. In the
wavelength range covered by our observations, and for the redshifts
of our clusters, the principal emission lines that are observable are
OII (3727 \AA), H-$\beta$ (4860 \AA) and the OIII (4959+5007 \AA)
doublet. Especially the combination of the latter two features
facilitated the recognition of the lines. Finally, fits were made to
the uncleaned, wavelength-calibrated spectra in order to determine
the wavelengths of the suspected emission-lines, and the implied
redshifts. From the comparison with the absorption-line redshifts
(see Section~4.2) it is found that the estimated errors in the
emission-line redshifts are between 40 km/s for redshifts based on
more than one emission line, and 80 km/s for redshifts based on a
single line.

In total, the 175 Optopus exposures (with multiple exposures of a
plate counted as one exposure) have yielded redshift estimates based
on the absorption-line spectrum for 5070 galaxies, and redshift
estimates based on emission lines for 1252 galaxies. For 360 galaxies,
independent observations (on purpose!) have yielded more than one
estimate of the absorption-line redshift, while for 47 galaxies at
least two independent estimates of the emission-line redshift were
obtained. For 666 galaxies, both an absorption- and an emission-line
redshift were estimated independently. In total we have obtained
for 5634 galaxies a redshift estimate, which corresponds to an average
of more than 32 galaxy redshifts per exposure.

In considering this average number of galaxy redshifts per Optopus
exposure, it should be noted that per exposure on average 1 of the
candidate galaxies turned out to be a star, while one or two fibres
per exposure were (purposely) positioned on blank sky. It is evident
that for spectroscopy of objects with the average surface density of
galaxies in rich clusters down to an integrated R-magnitude of about
17.5, the Optopus system with 50 fibres is ideally suited and very
efficient.

\section{The Reliability and Quality of the Redshift Estimates}

As our spectroscopic data will be used for several types of analysis,
either by itself or in combination with literature data, it is
important that we assess the quality and reliability of our redshift
estimates. We will discuss the following aspects of the reliability
and quality of our data. First, we will show that our redshift
estimates are not biased. Then we will discuss the reliability of
individual redshift estimates, which we derive from the multiple,
independent estimates within our dataset. These multiple measurements
are also used to ascertain that the error estimates that we quote for
the derived redshifts are indeed 1-sigma errors. Finally, we show that
there is good agreement between our data and the literature data that
is available for some of the clusters that we observed.

\subsection{The Redshift Scale}

Galaxy redshifts that are derived from cross-correlation of the
spectrum with a template galaxy spectrum are, in principle, prone to
biases which may result from imperfections in the wavelength
calibration of the galaxy and/or template spectrum. For the spectra
of the programme galaxies, we established the correctness of the
linearity and zero-point of the wavelength scale to within 0.2 \AA\
from lamp spectra and sky-lines. Possible errors in the wavelength
calibration of the M31 template spectrum that we used could also
produce biases in our redshift estimates that are difficult to detect
within our dataset.

However, we checked that possible non-linearity and zero-point errors
in the wavelength scale of the M31 template are negligible for all
practical purposes. First, we constructed a synthetic template
spectrum at a radial velocity of exactly 0 km/s. On a perfectly flat
continuum we superposed the absorption features that produce most of
the weight in the correlation analysis, viz.\ Ca-II, G-band,
Mg-Ib. The absorption lines were represented by Gaussians with a
dispersion of 1.4 \AA\ at the appropriate positions and with the
appropriate strengths. This synthetic template is considered superior
to the M31 template as regards the absence of wavelength calibration
errors, even though the absorption profiles are not perfect as far as
shape is concerned.  However, the synthetic template is clearly
inferior to that of the M31 nucleus as regards the correlation
strength for actual galaxy spectra, because all secondary absorption
features, which do contribute to the correlation amplitude, are absent
from the synthetic template.

Cross-correlation of the synthetic template with a spectrum of the
K-giant star HD 185781 (obtained by one of us with the 2.5-m Isaac
Newton Telescope at La Palma), yields a heliocentric radial
velocity of $-65$ km/s for the latter. Although this does not
agree very well with the value of $-80$ km/s given by Abt and Biggs
(1972), it is in excellent agreement with a more recent determination
by Mayor (priv. comm.), who found $-67\pm0.3$ km/s. Having confirmed
the correctness of the zero-point of the synthetic template, we then
cross-correlated the synthetic template with the M31 template. This
yielded a heliocentric radial velocity of $-306$ km/s for M31, which
is very close to the value of $-295\pm7$ km/s given by de Vaucouleurs
et al.\ (1991). This again confirms the correctness of the zero-point
of the synthetic spectrum, and shows that the wavelength scale of the
template spectrum is also correct. This latter conclusion is also
borne out by the data in Figure~3. Here we display the difference
between redshift estimates for 3065 of our programme galaxies, for
which correlation with both the M31 and the synthetic templates
yielded redshift estimates with a S/N-ratio $\ge$ 3. On the basis of
Figure~3 we estimate that any remaining deviation of our redshift
scale from the correct one amounts to less than 15 km/s per
$\Delta$z = 0.1.

\begin{figure}[t]
\vbox{}
\caption[]{The difference $\Delta z$ between the redshifts obtained
with the M31 template and with the synthetic template, as a function of
redshift. We show the 3065 galaxies for which the cross-correlations
with the M31 and the synthetic template both have a S/N-ratio $\ge$
3. From these data we conclude that the non-linearity of our redshift
scale is at most  15 km/s per $\Delta$z = 0.1. The dashed line
corresponds to the velocity of the M31 template}
\end{figure}

\subsection{The Reliability of the Redshifts}

There are two properties of a redshift estimate that are important.
First there is the uncertainty of the estimate, or rather: its
estimated error. Second, there is the reliability of the estimate,
i.e.\ the probability that a second, independent measurement will
produce the same result to within the limits set by the uncertainty
of both estimates. For absorption-line redshifts, the uncertainty of
the estimate follows from the curvature of the peak of the
cross-correlation function, in combination with the noise in the
correlation function. An empirical check of the correctness of the
estimated errors is discussed below.

\begin{figure*}
\vbox{}
\caption[]{\\
$a)$ The distribution of the absolute value of the velocity difference
(in units of the combined estimated errors) for 392 independent pairs
of multiple absorption-line redshifts. A Gaussian with dispersion of
unity and normalized to the total number of pairs is shown for
comparison.  $b)$ The distribution of the error estimates for the 3672
absorption-line redshifts with S/N $\ge$ 3. $c)$ The distribution of
the error estimates for the 1398 absorption-line redshifts with S/N
$<$ 3.}
\end{figure*}

The reliability of an absorption-line redshift is equal to 1.0 minus
the conditional probability that any of the peaks in the correlation
function, other than the accepted peak, is the true indicator of the
redshift of the galaxy. If the accepted peak is the highest one
(which is very frequently, but not always, the case), it can be shown
(see Tonry and Davis 1979) that the reliability of the derived
redshift is (very) small when the amplitude of the correlation peak
is low in comparison to the fluctuations in the correlation function.
The ratio between the amplitude of the peak and the RMS value of the
fluctuations is generally referred to as the S/N-ratio of the
correlation peak. The reliability as a function of S/N-ratio
approaches a Heaviside function, i.e.\ for high values of the
S/N-ratio the reliability is very close to 1, while in a rather
narrow range of S/N-ratio the reliability changes from essentially 0
to 1.

In the following, we have not attempted to derive from first
principles the detailed dependence of the reliability on the
S/N-ratio of the correlation peak. Instead, we use a schematic
representation in which we assume that above a certain limiting
S/N-ratio all reliabilities are equal to unity. Below this limit we
subsequently are interested only in the average reliability of the
available redshifts (which thus automatically involves the
distribution of the S/N-ratios below the limiting value). In Appendix
A we first use a set of two double exposures to derive a limiting
S/N-ratio of 3.0 above which the reliability is unity. Below this
value we find an average reliability of 0.6. It should be realized
that this latter number applies only when, in each and every
correlation function, the dominant peak is blindly accepted as the
indicator of the redshift. As explained before, it is also valid
{\sl only} for the particular S/N-ratio distribution in the two
double exposures.

The simultaneous visual inspection of the galaxy spectra, and their
correlation functions (as discussed above) has yielded the verdict
`highly improbable' for quite a few redshift estimates with S/N-ratio
$<$ 3.0, and such estimates have been rejected. In Appendix A we
derive an empirical estimate of the reliability of the redshift
estimates in the sample of 5070 absorption-line redshifts that were
{\em accepted} after visual inspection of the spectra and correlation
functions. We confirm (from multiple measurements) that the average
reliability of the estimates with S/N-ratio $\ge$ 3.0 is larger than
0.99. The average reliability of the {\em accepted} estimates with
S/N-ratio $<$ 3.0 turns out to be about 0.95.

The reliability of the emission-line redshifts has also been
estimated empirically from double measurements. As described in
Appendix A, it turns out that emission-line redshifts based on at
least two lines have a reliability that is larger than 0.95.
Redshifts based on only one emission line have an average reliability
of about 0.8 (but the uncertainty in this number is of order 0.1).

Finally we have used the (dis-)agreement between absorption- and
emission-line redshifts for the 666 galaxies for which we could
determine both, to improve our reliability estimates. As detailed in
Appendix A, we find reliabilities of 1.00 for S/N-ratio $\ge$ 3.0
absorption-line redshifts, of 0.95 for the accepted S/N-ratio $<$ 3.0
absorption-line redshifts, of 0.97 for multiple-line emission-line
redshifts and of 0.81 for single-line emission-line redshifts.

\subsection{The Construction of the Redshift Catalogue}

With the reliabilities derived in the previous paragraph in mind, we
have constructed our final redshift catalogues as follows.

First, we include all 4403 accepted galaxy redshifts based on {\em
absorption lines only}, as well as the 565 galaxy redshifts based
{\em only on one or more emission lines}. In the case of concordant
multiple measurements we have adopted the estimate with the highest
S/N-ratio (absorption lines), or from the largest number of
(emission) lines.

Second, for the 666 galaxies with {\em both absorption- and
emission-lines} we have determined the redshift as follows. For the
586 galaxies for which the two estimates agree to within 500 km/s
(with the large majority of the differences less than about 200 km/s),
the redshift is computed as the unweighted average of both estimates.
In the 80 cases where the estimates disagree, we generally accepted the
estimate in the category with the highest average reliability. This
means that all 57 discordant single-line emission-line redshifts were
ignored, as well as the 3 multiple-line emission-line redshifts that were
not concordant with S/N-ratio $\ge$ 3.0 absorption-line redshifts.
However, 18 of the 20 S/N-ratio $<$ 3.0 absorption-line redshifts that
were not concordant with multiple-line emission-line redshifts were ignored,
but 2 were accepted, because the emission-line redshift turned out to
be unacceptably low. This produces a catalogue of 5634 galaxy
redshifts with an estimated overall reliability of 0.98; i.e. we
expect not more than about 120 redshifts in our catalogue to be
erroneous.

\subsection{The Uncertainties in the Redshift Estimates}

The formal uncertainty of an absorption-line redshift estimate follows
directly from the curvature of the peak in the correlation function
and the noise (Tonry and Davis 1979). We have empirically checked the
statistical meaning of the error estimates using the multiple
absorption-line redshift estimates with S/N-ratio $\ge$ 3.0 for 265
galaxies, which form 392 independent pairs. In Fig.~4$a$ we show the
distribution of the absolute value of the velocity difference for
these 392 independent pairs expressed in units of the two error
estimates added in quadrature. It can be seen that the distribution is
indeed close to normal with a dispersion of unity, which confirms that
the error estimates are indeed 1-sigma errors. In Figure~4$b$ the
distribution of the error estimates for the 3672 redshifts with S/N
$\ge$ 3.0 is given, and in Figure~4$c$ the same is given for the 1398
redshifts with S/N $<$ 3.0.

The errors in the emission-line redshifts have been estimated as
follows. There are 47 pairs of twice-measured emission-line
redshifts. Of these, 30 are based on more than one line. For these
pairs the RMS value of the absolute velocity difference is only 57
km/s. There are 14 pairs having one estimate based on a single line
and another based on at least two lines. Two of these pairs are
discordant, but the RMS value of the absolute velocity difference of
the 9 concordant pairs is 129 km/s (or 95 km/s if the largest
difference of 297 km/s is ignored). Finally, all three pairs
consisting of two single-line redshifts are concordant. Their average
velocity difference is also 95 km/s. We conclude that redshifts based
on more than one line on average have very small estimated errors of
about 40 km/s and are thus comparable to our `best' absorption-line
redshifts. Single-line redshifts are probably a factor of two less
accurate.

\begin{table}
\caption[]{
Comparison between ENACS and literature redshifts, for galaxies with
position differences of not more than 20$\arcsec$.}
\begin{flushleft}
\begin{tabular}{lccrrl}
\hline\hline
\noalign{\smallskip}
Cluster & \multicolumn{2}{c}{\# galaxies} & $\avrg{\Delta V}$ &
$\sigma_{\Delta V}$ & ref.\\
        & total & $|\Delta V| < 500$ & km/s & km/s & \\
\noalign{\smallskip}
\hline
\noalign{\smallskip}
A0119 & 59 & 41 & -176 & 152 &  1     \\
A0151 & 34 & 28 &  -96 & 193 &  2     \\
A0168 & 10 &  8 &   13 &  58 &  3,4   \\
A0548 & 29 & 27 &  -57 &  90 &  5     \\
A0754 & 29 & 23 &   25 & 103 &  5,6   \\
A0957 & 21 & 19 &   27 & 159 &  7     \\
A1069 &  6 &  5 &   13 & 117 &  7     \\
A1809 & 25 & 25 &   38 & 108 &  8,9   \\
A2052 & 22 & 22 &  -17 & 121 &  9     \\
A2717 & 17 & 17 & -105 &  95 &  10    \\
A3112 & 15 & 13 &  -58 & 215 &  11    \\
A3128 & 16 & 16 &   17 &  72 &  10    \\
A3158 & 16 & 14 &  -48 & 215 &  12,13 \\
A3558 & 36 & 32 &   15 &  99 &  14    \\
A3667 & 64 & 51 &   69 & 155 &  15    \\
 & &\hrulefill&\hrulefill&\hrulefill& \\
total &   & 341 &  -23 & 130 &        \\
\noalign{\smallskip}
\hline\hline
\end{tabular}
\end{flushleft}
{References:
1. Fabricant et al.\ (1993); 2. Proust et al.\ (1992); 3. Zabludoff
et al.\ (1993); 4. Faber \& Dressler (1977); 5. Dressler \& Shectman
(1988); 6. Zabludoff et al.\ (1990);
7. Beers et al.\ 1991; 8. Hill \& Oegerle (1993);
9. Malumuth et al.\ (1992); 10. Colless \& Hewett (1987);
11. Materne \& Hopp (1983);
12. Chincarini et al.\ (1981); 13. Lucey et al.\ (1983);
 14. Teague et al.\ (1990);
15. Sodr\'e et al.\ (1992)}
\end{table}

\subsection{Comparison with Redshifts from the Literature}

For 15 of the clusters observed by us (viz. A119, A151, A168,
A548, A754, A957, A1069, A1809, A2052, A2717, A3112, A3128, A3158,
A3528, A3558 and A3667), galaxy redshifts have been published by
other authors. We have cross-identified the galaxies on the basis of
positional and velocity agreement, requiring that the position
difference be less than 20$\arcsec$ and the velocity difference less
than 500 km/s. The results are summarized in Table~2.  We conclude
that the agreement between our redshifts and those in the literature
is very good. We do not find systematic velocity offsets at a level
which would prevent a useful combination between our data and those
from the literature.

\section{The Calibration of the Photographic Photometry}

Wide-field photographic imaging with Schmidt telescopes is currently
the fastest and only practical way to cover large areas of sky, and
the POSS and SERC surveys together provide coverage of all our
clusters. All photometry for the galaxies in our programme is based on
these surveys, and was obtained during the construction of the galaxy
catalogues. In order to establish the zero-points of the photographic
photometry we have obtained CCD photometry for about half of our
clusters.

\begin{table*}
\caption[]{Some details of the CCD photometry.}
\begin{flushleft}
\begin{tabular}{lllrl}
\hline\hline
\noalign{\smallskip}
\multicolumn{2}{l}{Observing period} & Tel. & Filters & Clusters \\
\noalign{\smallskip}
\hline
\noalign{\smallskip}
01/11 $-$ 06/11 & 1989 & 1.54-m & B,R & A2717, A3009, A3108, A3141, A3158,
A3223, A3795, A3822 \\
25/04 $-$ 28/04 & 1990 & 1.54-m & B,R & A0754, A0957, A0978, A1069, A1399,
A1809, A2029, A2040, A2048, A3528, \\
                &      &        &     & A3558, A3559, A3562 \\
17/10 $-$ 18/10 & 1990 & 1.54-m & B,R & A0151, A2426, A3094, A3667, A3864 \\
08/10 $-$ 10/10 & 1991 & 0.92-m &   R & A0168, A0367, A2383, A2480, A2764,
A2819, A2915, A3112, A3122, A3264, \\
                &      &        &     & A3651, A3691, A3799, A3809 \\
24/11 $-$ 29/11 & 1992 & 0.92-m &   R & A0295, A0514, A0548, A2734, A3128,
A3142, A3341, A3825 \\
\noalign{\smallskip}
\hline\hline
\end{tabular}
\end{flushleft}
\end{table*}

Galaxy photometry is notoriously difficult, because there is no such
thing as {\em the} magnitude of a galaxy. Basically the problem is
that one needs to define an aperture within which the integrated
brightness of the galaxy is determined. Due to the large variations in
the brightness profiles of galaxies (both in slope and in
characteristic scale), any aperture one cares to define (be it metric
or isophotal) has a different physical meaning for different
galaxies. In short: it is impossible to derive from the brightness
distribution one single number that can be used in the same manner for
all galaxies, without at the same time giving a pictorial description
of the individual galaxies. This basic problem is very much in
evidence in the calibration of the photographic photometry.

\subsection{The Photographic Photometry}

During the construction of the cluster galaxy catalogues, several
object parameters were obtained from the plate-scanning with
the {\sl Astroscan} measuring machine (see Section 2.2).  We used a
detection threshold above sky of between 5 and 7 times the noise in
the sky background.  For the photographic photometry, this threshold
also served as a limiting isophote. For each galaxy, the photometric
parameter $P_{\rm phot}$ was derived as the square root of the sum of
the photographic densities above sky within the isophote defined by
the detection threshold.

For the 103a-E emulsion of the red PSS survey, this threshold thus
corresponds to an isophote (above sky) of about 40\% of sky. This
number is based on a noise in photographic density of 0.05 per 500
$\mu$m$^2$ (the effective size of a resolution element) and
$\gamma\simeq 2$ (the slope of the characteristic curve of the
emulsion). For the IIIa-J emulsion of the green SERC survey, with
corresponding values of 0.04 and 3.5, the detection threshold
corresponds to an isophote (above sky) of about 20\% of sky. The
nominal values of the sky brightness are about 20 and 22
mag/arcsec$^2$ for the 103a-E and IIIa-J emulsions respectively.
Hence, the limiting isophotes of our photographic photometry would be
expected to be
of the order of 21.0 and 23.5 mag/arcsec$^2$ on the red and green
plates respectively. However, actual sky brightnesses are likely to
be higher than these nominal, minimum values, probably by as much as
1.0 mag. This assumption is consistent with the practical plate
limits for the detection of stellar objects on these plates of about
20.5 and 22.5 mag respectively. The actual values of the limiting
isophotes will vary somewhat between clusters, primarily because of
differences in sky brightness between plates, which are of the order
of 0.2 mag (Dalton et al.\ 1992).

\subsection{The CCD photometry}

In order to calibrate the photographic photometry, we have obtained
CCD-images of subsets of the galaxies in about half of our clusters.
We used the 1.54-m (`Danish') and 0.92-m (`Dutch') telescopes at La
Silla. In Table~3, we give a summary of the dates of the
observations, the telescopes and filters used, and the clusters for
which we obtained data. The weather was not always cooperative, and
some of the data obtained are not of photometric quality. There are 4
clusters with very large internal spreads (of between 0.50 and 0.80
mag) with respect to the average calibration relation discussed
below. We have excluded these clusters from the calibration, on the
reasonable assumption that the large internal dispersions are due to
non-photometric conditions.

For all galaxies in the CCD-frames for which photographic photometry
was available, an isophotal R-magnitude within the 25 mag/arcsec$^2$
R-isophote ($R_{25}$) was determined (see e.g.\ Le F\`evre et al.\
1986).  For the galaxies that had B-band CCD-data available as well,
the B-band isophotal magnitude within the 25 mag/arcsec$^2$
B-isophote ($B_{25}$) as well as the B-R colour within the 25
mag/arcsec$^2$ R-isophote $(B-R)_{25}$ were also determined. One
might argue that a much brighter isophote should have been chosen, in
order to adhere more closely to the apertures in the photographic
photometry. That would certainly have decreased the influence of the
shapes of the individual galaxy brightness profiles on the
photometric calibration, although it would be very difficult to
achieve identical apertures for all objects in the two sets of
photometry. By adopting $R_{25}$ we have wilfully accepted a larger
influence of the variety of the shapes of galaxy brightness profiles
in our calibration, but we have ensured uniformity and compatibility
with many programs of galaxy photometry in the literature.

\subsection{The Quality of the Calibration}

\begin{figure*}
\vbox{}
\caption[]{The calibration relations between the photometric parameter
$P_{\rm phot}$ and the $R_{25}$ isophotal CCD magnitude for \\
$a)$ the 103a-E emulsion of the red PSS survey and $b)$ the IIIa-J
emulsion of the green SERC survey. Note that the individual
zero-points of the clusters have been taken into account.}
\end{figure*}

In Figure~5 we show the relations between the photographic magnitude,
i.e.\ $P_{\rm phot}$, and the isophotal R-magnitude ($R_{25}$)
derived from the CCD-imaging, for 100 galaxies in 11 clusters scanned
on PSS plates (red) and for 314 galaxies in 28 clusters scanned on
SERC IIIa-J plates (green). For $R_{25}$ between about 14 and 18, the
relation between the logarithm of $P_{\rm phot}$ and $R_{25}$ appears
indeed to be quite linear, with a slope not very different from 5.0
(the zero-order expectation). In producing the calibration relations
in Figure~5, we have taken out differences in the zero-points for
individual clusters, which are to be expected as a result of the
variations in limiting isophote. As fits to the individual
calibration relations do not reveal significant differences in slope,
we determined the zero-points from a maximum-likelihood fit to the
calibration data that solves simultaneously for a universal slope and
individual zeropoints per cluster. The global dispersions with
respect to the average relations in Figure~5 are 0.24 and 0.34 mag
for the PSS and SERC calibrations respectively.

The dispersion in the global calibration relations consists of the
following contributions. First, there are random measuring errors in
the photometry. Second, there is a contribution due to the
differences in limiting isophotes, which makes itself felt through
the large variation in brightness profiles of galaxies. As a result,
the correlation between the two isophotal magnitudes is far from
perfect. In a calibration dataset like ours one thus expects a
systematic dependence of the difference between the two magnitude
estimates on the ratio of the isophotal apertures, as that ratio
reflects the `slope' of the brightness distribution.  Finally, there
is an extra contribution to the spread in the SERC calibration
relation from the appreciable range in galaxy colours, which is a
result of the fact that we compare red CCD magnitudes to green
photographic magnitudes.

Both the aperture and the colour effect are clearly detectable in our
calibration data. The aperture effect is, to first order,
proportional to the logarithm of the ratio of the two apertures,
log($A_{\rm CCD}/A_{\rm phot}$). The global slope of the calibration
relation, the individual
zero-points of the clusters and the coefficient of the
aperture ratio were all obtained from a combined maximum-likelihood
fit to the relation between $R_{25}$ and $P_{\rm phot}$ for all
galaxies in the sets of clusters that were scanned on the same
emulsion. We find that the size of the aperture effect is about 0.5
mag per decade in aperture ratio for the PSS data and about 1.0 mag
per decade for the SERC data.  The factor of two difference between
the two constants is due to the difference in contrast index of the
two emulsions. The colour effect was found, from a limited subset of
clusters with $(B-R)$-data, to amount to about 40\% of the measured
$(B-R)_{25}$ colour. This seems reasonable since the SERC IIIa-J
passband is quite a bit redder than $B$, while for red galaxies the
effective wavelength is even redder than average.

\begin{table}
\caption[]{Zero-points and internal dispersions of the photometry for
clusters with useful calibration data.}
\begin{flushleft}
\begin{tabular}{lrrrl}
\hline\hline
\noalign{\smallskip}
  Cluster & $\Delta$ zero & $\sigma_{\rm mag}$ & $N_{gal}$ & colour \\
\noalign{\smallskip}
\hline
\noalign{\smallskip}
  A0151       &  0.04 & 0.28 & 15 & R \\
  A0168       & -0.03 & 0.14 &  3 & R \\
  A0295       & -0.14 & 0.16 &  8 & R \\
  A0367       & -0.17 & 0.15 & 16 & G \\
  A0514       & -0.05 & 0.37 & 25 & G \\
  A0754       & -0.52 & 0.20 & 13 & R \\
  A0957       & -0.14 & 0.16 &  8 & R \\
  A0978       &  0.06 & 0.18 &  8 & R \\
  A1069       &  0.40 & 0.20 & 11 & R \\
  A1809       &  0.14 & 0.36 &  9 & R \\
  A2040       & -0.02 & 0.32 & 13 & R \\
  A2048       &  0.03 & 0.28 & 16 & R \\
  A2383       & -0.21 & 0.47 & 10 & G \\
  A2426       &  0.20 & 0.35 &  4 & R \\
  A2480       & -0.15 & 0.15 &  4 & G \\
  A2717       & -0.27 & 0.49 &  6 & G \\
  A2734       &  0.56 & 0.32 &  5 & G \\
  A2764       &  0.05 & 0.35 &  7 & G \\
  A2819       & -0.09 & 0.30 & 38 & G \\
  A2915       &  0.28 & 0.49 &  6 & G \\
  A3009       &  0.45 & 0.40 &  4 & G \\
  A3094       &  0.51 & 0.33 &  5 & G \\
  A3108       &  0.30 & 0.41 &  6 & G \\
  A3112       & -0.23 & 0.38 & 13 & G \\
  A3122       &  0.29 & 0.34 & 10 & G \\
  A3128       & -0.12 & 0.34 & 28 & G \\
  A3141       & -0.29 & 0.29 &  3 & G \\
  A3142       &  0.42 & 0.40 &  6 & G \\
  A3158       & -0.12 & 0.25 & 14 & G \\
  A3223       &  0.13 & 0.40 & 14 & G \\
  A3264       &  0.20 & 0.45 & 11 & G \\
  A3528       & -0.51 & 0.22 & 16 & G \\
  A3558+59+62 &  1.69 & 0.44 & 13 & G \\ 
  A3651       & -0.26 & 0.41 & 10 & G \\
  A3667       & -0.39 & 0.48 & 27 & G \\
  A3691       & -0.16 & 0.31 & 14 & G \\
  A3795       & -0.18 &      &  2 & G \\ 
  A3809       &  0.03 & 0.47 &  6 & G \\
  A3822       & -0.01 & 0.35 &  4 & G \\
  A3864       & -0.02 & 0.30 &  4 & G \\
\noalign{\smallskip}
\hline\hline
\end{tabular}
\end{flushleft}
\end{table}

For the subset of galaxies for which we have information on the
aperture-ratio and on the $(B-R)$ colour we find that, when we
correct for these effects by reducing all $R_{25}$ values to a
reference aperture-ratio and, where possible, to a reference $(B-R)$
colour, the global dispersion in the calibration relation indeed
decreases, to a value of about 0.20 mag. This still includes a
contribution from the variation in the `slope' of the brightness
distributions, and is thus consistent with estimates of the combined
random noise in the photometry of about 0.15 mag.

For the large majority of the galaxies for which we obtained
redshifts we have no information on aperture-ratio or colour. To convert
photographic magnitudes into $R_{25}$'s, we can thus do no better than
apply the average calibration relations shown in Figure~5 for these
galaxies. For the 39 clusters for which we do have usable
calibration data, and for which we could therefore estimate an individual
zero-point, we applied the latter, even if it did not differ
significantly from the average value. In Table~4 we show the values
of the zero-point offset (individual cluster value minus average) as
well as the internal dispersion for each
of these 39 clusters. The Table gives an indication of the maximum
zero-point errors that we are likely to have made for the clusters
without calibration data.

When using our calibrated R-band photometry, the following points
must be realized. First, within a cluster for which no individual
zero-point is available, the relative photometry is not affected. In
other words: for such clusters one can still study the luminosity
distribution. However, when combining photometric data from several
clusters, one has to know whether the zero-point of the photometry
was estimated for the given cluster, or whether it was assumed to be
equal to the average value. From Table~4 we estimate that the RMS
value of the difference between the assumed and the actual zero-point
for a cluster without individual calibration is 0.27 mag. For the
clusters with calibration, one finds from Table~4 that the quality of
the individual zero-points is of the order of 0.05 to 0.1 mag.

\begin{table}[htb]
\caption[]{
Comparison between ENACS and literature magnitudes, for galaxies
whose positions agree to within 20$\arcsec$.}
\begin{flushleft}
\begin{tabular}{lrlrrl}
\hline\hline
\noalign{\smallskip}
Cluster & \# gal. & band & $\avrg{\Delta {\rm mag}}$ & $\sigma_{\Delta {\rm
mag}}$ & ref.\\
\noalign{\smallskip}
\hline
\noalign{\smallskip}
A0119 & 59 & R        &  0.39 & 0.10 & 1 \\
A0754 & 33 & R        & -0.26 & 0.08 & 2 \\
A2717 & 17 & B$_J$    & -0.98 & 0.11 & 3 \\
A3128 & 16 & B$_J$    & -1.67 & 0.11 & 3 \\
A3667 & 91 & B$_{25}$ & -1.70 & 0.22 & 4 \\
\noalign{\smallskip}
\hline\hline
\end{tabular}
\end{flushleft}
{References:
1. Fabricant et al.\ (1993); 2. Fabricant et al.\ (1986);
3. Colless (1989); 4. Sodr\'e et al.\ (1992)}
\end{table}

\subsection{Comparison with Photometry from the Literature}

For 8 clusters in our sample there are also magnitudes available in the
literature. In 3 cases, viz.\ A0151, A0548 and A2052, the magnitudes
were estimated by eye, and we did not use those data to check the
quality of our photometry. For the remaining 5 clusters we show, in
Table~5, the results of a comparison between our photometry and that
from the literature. It must be realized that for A0754, A2717, A3128
and A3667 the literature data also consist of photographic
photometry calibrated by CCD-imaging. For A119, the photometry is
based on CCD-imaging of individual galaxies, at least for the
brighter galaxies that carry essentially all of the weight in the
comparison. Galaxies were cross-identified on the basis of
positional agreement only; we required that the positional difference
be less than 20$\arcsec$. From Table~5 we conclude that the offsets
$\Delta$mag between our data and the literature data are consistent
with a spread of about 0.3 mag in the zero-points of our magnitudes.

The agreement with the photometry from the literature (as judged from
the dispersion in $\Delta$mag) is better than one would expect from
the dispersions $\sigma_{\rm mag}$ in our calibration relations, or
from those reported by the other authors, for A119, A754, A2717 and
A3128.  This is not surprising because in those four cases we compare
our photographic magnitudes directly to photographic magnitudes based
on the same plate material. The dispersion in $\Delta$mag is then
primarily due to the (unknown) differences in threshold in the two
sets of photographic photometry (i.e. the aperture effect), and to
the colour differences between galaxies. The larger dispersion in
$\Delta$mag for A3667 is most likely due to the fact that the
photographic magnitudes were converted from $B_J$ to the isophotal
$B_{25}$ system, which probably introduces an additional aperture
term.

\begin{figure}[t]
\vbox{}
\caption[]{The distribution of the R$_{25}$ magnitudes of all 5615
galaxies.}
\end{figure}

\subsection{The Completeness of the Redshift Surveys}

In Figure~6 we show the combined apparent magnitude distribution for
all 5615 galaxies for which we obtained a redshift (for technical
reasons photographic photometry is lacking for 19 galaxies).

As explained in Section 2.2, we have attempted to measure a redshift
for each of the 50 brightest galaxies in the region of each Optopus
plate. That means that we obtained spectra for essentially all
galaxies above a given limiting R$_{25}$ magnitude, which for the
different clusters varies between 17 and 18. However, not all spectra
have yielded a redshift. There are several reasons for that, the most
important ones of which are the following. First, an isophotal
magnitude brighter than a given limit does not always guarantee a
sufficiently bright central surface brightness. This seems to be true
in particular for the brightest galaxies, which are quite large
compared to the size of the fibres. As a result, a redshift was
obtained on average only for 2.2 of the brightest 5 galaxies (and for
2.5 of the 5 next brightest galaxies). Variations between galaxies in
the strengths of the absorption and emission lines also contribute to
a less-than-100\% score. At the faint end, i.e. near the magnitude
limit this seems to be the main reason for lack of success.

We have calculated the completeness (i.e. the ratio of the number of
redshifts obtained over the number of galaxies observed) for each
cluster as a function of magnitude. The completeness increases in
general from the brightest to intermediate magnitudes and then either
stays constant or decreases somewhat towards the magnitude limit.
Most clusters have a 'maximum' completeness at R$_{25}$ = $17.1 \pm
0.3$. For 52 clusters this maximum completeness is between 0.7 and
0.9. For 30 clusters it is between 0.6 and 0.7, and for the remaining
25 clusters it is below 0.6. Among these 25 are the 5 clusters that
were already noted in Section 3.1 to have been observed in very bad
conditions. Frequently, several redshifts are available beyond the
magnitude at which the completeness is maximum, and therefore the
maximally complete samples contain in total about 500 galaxies less
than the 5634 for which we have measured redshifts.

Because the information on the completeness is relevant for some
types of analysis, we will give details about it when we make
available the full catalogues.

\section{The Identification of Systems in Redshift Space}

\begin{figure*}
\vbox{}
\caption[]{
 Distribution of radial velocities in the directions of the 107 target
 ACO clusters studied in the ESO Nearby Abell Cluster Survey (ENACS).
 Solid bars indicate velocities derived solely from absorption lines,
 dashed bars indicate velocities derived from emission lines or from
 emission and absorption lines. The total number of galaxy redshifts
 in a survey is shown next to the Abell number of the cluster.}
\label{f:velbar}
\end{figure*}

As we are interested in studying the properties of rich clusters of
galaxies, our first concern is to define the physically relevant
systems in our redshift surveys by identifying superimposed fore- and
background galaxies. This problem is illustrated in Figure~7, in
which we present for each of our 107 target ACO clusters (lines of
sight hereafter) the radial velocities in a `bar-plot'.

Several methods have been proposed in the literature for identifying
systems in redshift surveys in a more or less objective manner. In
all existing methods some assumptions have to be made about the
systems that one wants to detect, or some parameter has to be chosen
which influences in a fundamental way the detection of systems or
their properties. Because, as far as we know, there does not exist a
method that is free from interpretation, we try to use a method that
does not produce results that are clearly in conflict with either the
qualitative visual impression from Figure~7 or with well-established
properties of clusters.

In the context of the present discussion we are only interested in
defining the overall properties of the systems, i.e.\ their
existence, their average redshift, and their approximate extent in
velocity space.  Therefore, a good method should on the one hand be
able to separate the obvious physical systems from fore- and
background galaxies. In addition it should separate systems if there
is more than one in a pencil-beam survey. Finally, it should not
break up physical systems into subsystems.

For many of the larger systems, consisting of several tens of
galaxies, the overall properties will depend only weakly on the
details of the method used. The main reason for this is that
discretization is not important for such systems. However, it is
probably unavoidable that the properties of the smallest system that
the eye detects in Figure~7 will depend rather strongly on the
details of the employed method.

The velocity bar-plots in Figure~7 show that the density contrast
between the systems and the fore- and background galaxies is rather
sharp. This quite naturally leads to two types of solution to the
problem of system definition; one that uses the high density and
compactness of the systems, and one that is based on the emptiness,
the `gaps' in velocity, between the systems.  In the latter case, if
two adjacent galaxies in the velocity distribution are to belong to
the same group, their velocity difference should not exceed a certain
value, the {\em velocity gap}.

An example of the first type of solution is the method proposed by
Pisani (1993).  In this method a cluster is defined as a single peak
in the probability density that underlies the distribution of
galaxies along the line of sight. The method is non-parametric, as it
does not require an input parameter, such as a limiting gapsize.
However, the method employs a limiting probability for assigning
individual galaxies to a given system, which influences the
properties of the resulting systems. As White (1991) illustrates with
N-body models, clusters that are spatially compact do not necessarily
show a single peak in velocity space. Therefore, the basic assumption
of the method is probably not true for at least some of the systems.
We have applied the method to our data with the probability limit
proposed by Pisani, and found that it breaks up systems which we
consider compact into smaller sub-systems. As we do not want to
prejudice the separation of systems into possible sub-systems at this
stage, we have decided not to use this method.

For a given physical system the distribution of gap sizes evidently
depends on the number of velocities sampled, and on the velocity
width of the system. To ensure uniformity in the definition of
systems, the limiting gapsize should therefore, at least in
principle, take into account the number of velocities measured for a
given system as well as the velocity dispersion. However, because
these are exactly the properties we are trying to define for a given
system, the limiting gapsize must be estimated from the velocity
width and population of the well-sampled systems, of which the global
definition is not problematic.

Zabludoff et al.\ (ZHG 1990) propose a two-step scheme along these
lines, in which first a fixed gap of 2000 km/s is applied to identify
the main systems.  Subsequently, a gap equal to the velocity
dispersion $\sigma_V$ of the system is applied, to eliminate outlying
galaxies.  In this case, the gap of 2000 km/s is based on the overall
velocity widths of known clusters, while the second step employs the
detailed information in the system that one is defining. We have used
this method on our data, but find that we need to use a different
choice of the parameters, in order to avoid the merging of separate
systems into larger units (as happens e.g.\ in the case of A0151 and
A2819).  Because the ZHG method has been applied to find systems in a
comparable survey in the north, we will discuss it in some more
detail below.

Another possible method considers gaps as the result of a Poissonian
process, with the expectation value of the gapsize chosen equal to
the median gapsize. The motivation for this method is again that it
does not require a physical input parameter, only the specification
of a limiting probability. Large gaps with a probability smaller than
the specified limit define the physical systems. The method works
well for the well-sampled surveys and the results then do not depend
critically on the choice of the limiting probability. However, for
lines of sight with less than about 30 redshifts, most gapsizes are
close to the median value and the method is no longer sufficiently
discriminative.

\begin{figure}
\vbox{}
\caption[]{The distribution of the velocity differences between
galaxies (within a survey) that are adjacent in redshift,
summed over all 107 surveys.}
\end{figure}

Ideally, we want to use the information contained in our redshift
surveys, about the well-sampled as well as the less well-sampled
systems. We also want to avoid making assumptions about e.g.\ the
velocity dispersions of the systems in our surveys.  Therefore, we
have determined for each survey the distribution of the velocity gaps
between galaxies adjacent in redshift, and summed these 107
distributions. The result is shown in Figure~8. It is clear that for
small gaps the distribution is determined by the internal velocity
structure of the real cluster systems. The very large gaps, which
occur between galaxies that are not associated with the systems, are
more or less uniformly distributed. The optimal choice for the
definition of systems is therefore the smallest gap value for which
the distribution in Figure~8 is still flat. For such a gap, galaxies
that are associated with the systems are not separated, while most of
the field galaxies will not be linked to the systems.  On the basis
of the distribution in Figure~8, we adopt a value of 1000 km/s for
the definition of the physically relevant systems in our survey.

In Table~6 (at the end of this paper) we list the properties of the 220
systems with at least 4 members, identified using a 1000 km/s fixed
gap, in the 107 lines of sight observed in our ESO Nearby Abell
Cluster Survey. The choice of the lower limit of 4 members is
motivated by a comparison between the systems defined with a fixed
gap of 1000 km/s and those defined with the method proposed by ZHG.
With the latter method all main systems in Table~6 are also found,
with essentially the same average redshifts, except for the two
systems with a velocity difference of about 4000 km/s in A151 and the
two systems in A2819 with a similar velocity difference.  The system
definition according to ZHG merges both sets of systems which, on the
basis of the redshift histograms, we do not consider acceptable.

For the smaller systems, in particular those with N$\la$10, the
merging of separate groups in our list into single systems by the ZHG
method is a fairly common phenomenon. In some cases, the elimination
of outlying galaxies on the basis of the provisional value of the
velocity dispersion breaks up systems (defined with our method) in
two sub-systems, while in a few other cases our method does not
define a (small) system while the ZHG method does. In general, these
effects confirm that the definition of small systems (say, with $\le$
10 members) is influenced by noise and by the details of the method.
Most of the N$<$4 systems that are defined with our method are not
found by the ZHG method. In Table~6 we have indicated on which
N$\ge$4 systems our fixed-gap method and the ZHG method disagree.

\begin{figure*}
\vbox{}
\caption[]{The distribution on the sky of the 96 lines of sight in
ENACS that have $b\le -10\degr$. Dashed lines are at constant values
of galactic latitude, the two dotted lines are at constant
declination. The symbols indicate the lines of sight and their size
reflects the number of galaxies found in the main system in the
corresponding redshift survey.}
\end{figure*}

\begin{figure*}
\vbox{}
\caption[]{The projected distribution of galaxies with a measured
redshift, in the 28 `structure' clusters for which more than one
Optopus aperture plate was used. The size of the symbols is an
indication of the brightness of the galaxies. The median position
of the galaxies is indicated by a cross, and listed as well. The diameter
of the Optopus plates is 33$\arcmin$.}
\end{figure*}

Note that no systems were found in the direction of the cluster
candidates A2502 and A3144; however, both were observed in very poor
conditions. Note also that we have not listed velocity dispersions.
This is because the system definitions in Table~6 are based only on
the redshift distributions. Because it is not certain at this stage
that each and every galaxy within the redshift limits of a system is
indeed a member of that system, one must discuss the plausibility of
the membership of each galaxy on the basis of velocity {\em and}
position information, to obtain a dynamically meaningful estimate of
the global velocity dispersion (see e.g.\ den Hartog and Katgert 1995
and Mazure et al.\ 1995). We refer to the latter for a determination
of the distribution of global velocity dispersions, based on the data
presented here in combination with data from the literature.

\section{Field Contamination and Superposition Effects}

Abell (1958) and Abell et al.\ (1989) identified clusters, from visual
inspection of survey plates, as overdensities in a 2-dimensional
projection of a 3-dimensional galaxy distribution. It is evident from
Figure~7 that a very large fraction of the rich and nearby Abell
cluster candidates that we studied are coherent structures in
velocity space. However, at the same time, the bar-plots also clearly
show the importance of superposition effects, which certainly will
have influenced to some extent the observed 2-dimensional
characteristics of clusters, such as richness or morphology.

The amount of field contamination in our survey can be quantified as
follows. About 75\% of the 5634 galaxies are in the largest system
found in the direction of each of the rich Abell cluster candidates.
Note that this number refers to the galaxies within $\approx 1$ \Mpc
of the cluster centre and down to $R \approx 17$, and that it applies
to clusters with $z \la 0.1$. Field contamination is thus found to be
quite substantial for such clusters. However, about half of the 1422
galaxies that are outside the main systems are not in the field
either, but in a secondary system. Field contamination in the strict
sense of the word therefore probably amounts to at most 12\% in our
survey.

{}From the data in Table~6 we have calculated the fraction of redshifts
in each pencil beam that is contained in the main system. In 85 out
of 103 cases (we ignored the 4 pencil beams with less than 10 measured
redshifts), i.e.\ in 83\% of the cases, the main system is found to
contain at least half of the total number of measured redshifts. It
is of interest to compare this with an earlier estimate by Lucey
(1983). Using models of the galaxy distribution, Lucey concluded that
between 15 and 25 per cent of the clusters in the Abell catalogue
have a true galaxy population that is less than half the apparent
population of the cluster. Although our result and Lucey's estimate
refer to slightly different apertures and to different redshift
ranges, the agreement is quite satisfactory. In our data there is a
clear dependence of this fraction on the total number of measured
redshifts in the pencil beam. Of the 31 pencil beams with at least 46
measured redshifts, 29 (i.e.\ 94\%) have a main system with at least
half of the total number of redshifts; on the contrary: of the 35
pencil beams with less than 35 redshifts, 24 (i.e.\ 69\%) have a
main system with at least half the total number of redshifts. As the
number of redshifts that we obtained correlates more or less with
richness, this trend is not unexpected.

Field contamination will thus have influenced the richness of some of
the clusters appreciably. Any cluster sample complete with respect to
apparent richness will contain some clusters that should not be in
the sample (those with higher-than-average contamination), while some
clusters that should be in the sample will not have been included as
a result of lower-than-average contamination. However, Lucey (ibid.)
finds that contamination modifies the richness distribution of a
cluster sample only slightly. In our discussion of the distribution
of velocity dispersions for a complete, volume-limited cluster sample
(Mazure et al.\ 1995) we study the effect of field contamination on
richness, and in particular its effect near the richness limit of the
sample.

A related question concerns the number of spurious rich clusters that
result from the superposition of two poorer systems. As there is no
unique definition of a spurious cluster, we will consider an Abell
cluster candidate to be spurious if at least two redshift systems are
found, with the number of redshifts in the main system exceeding that
in the next richest system by not more than a factor of two. We have
estimated the fraction of such spurious clusters in our sample as
follows. In 18 out of 103 cases we find a secondary system with at
least half as many galaxies as in the main system. However, this is
very likely to be an overestimate of the true fraction of spurious
clusters, for the following reason. The large majority (13 out of 18)
of these cases occur when the number of redshifts in the secondary
system is less than about 10. It is thus likely that small number
statistics has artificially raised the number of spurious clusters.
If we limit ourselves to the 64 cases in which the total number of
redshifts in the main and (if present) secondary system is at least
30, we find that only 5 (i.e. 8\%) of our clusters are spurious
according to the above definition. If, alternatively, we base our
estimate on the 79 cases in which the number of redshifts in the main
system exceeds 15, we find that 10 \% of the clusters is spurious.

Although it is thus not completely straightforward to make a statement
about the number of spurious clusters in our sample, we conclude that
probably only in about 10\% of the cases an $R\ge 1$ , $z \le 0.1$
entry in the ACO catalogue is the result of a superposition of two
almost equally rich (but relatively poor) systems. This would imply
that 8 of the 10 apparently spurious clusters among the 24 clusters
with $\le 15$ redshifts in the main system must be attributed to small
number statistics, which does not seem unreasonable.

Our conclusion that about 90\% of the rich and relatively nearby ACO
cluster candidates that we studied appear to be real rich clusters
might, at first sight, seem to be in conflict with the result of
Sutherland (1988). From a comparison of the angular and radial
correlation functions of Abell clusters, Sutherland concluded that
superposition must play an important r\^ole. However, it must be
realized that his analysis was based on a sample which included many
clusters of richness 0, for which superposition presumably is much
more serious than for our $R\ge 1$, $z \le 0.1$ clusters. Therefore,
we do not consider the result of Sutherland to be in conflict with
our data.

\section{The Spatial Distribution of Clusters and Galaxies}

In Figure~9 we show the distribution on the sky of the 96 pencil-beam
surveys for which $b\le -10\degr$.  The size of the symbols reflects
the number of galaxies found in each of the main systems found in these
surveys. The dotted lines indicate the declination limits of the
ENACS. In a companion paper (Mazure et al.\ 1995) we will discuss the
completeness of the set of clusters that one obtains by combining the
main systems in Figure~9 with data from the literature.

There are a few remarkable concentrations of clusters to be seen in
Figure~9. Some of these are probably due to chance superposition but
a few correspond to well-known superclusters. The concentration in the
region $245 \degr \le l \le 265 \degr$, $b \approx - 54 \degr$ is part
of the Horologium-Reticulum supercluster (Lucey et al.\ 1983), and
contains A3093, A3108, A3112, A3122, A3128, A3158, and possibly
A3202. Three other concentrations, viz.\ those at $l \approx 230
\degr$, $b \approx - 25 \degr$ (containing A548 and A3341), at $l
\approx 7 \degr$, $b \approx - 35 \degr$ (containing A3682, A3691,
A3693, A3695, A3696 and A3705), and at $l \approx 338 \degr$, $b
\approx - 46 \degr$ (containing A3806, A3822 and A3825) were already
detected by Zucca et al.\ (1993),  on the basis of more limited
redshift information.

In Figure~10 we show for the 28 (`structure') clusters, for which
several Optopus fields were defined, the projected distribution of
the galaxies for which we obtained redshifts. The size of the symbols
indicates the apparent brightness of the galaxies, while the outer
contours delineate the outer boundary of the area covered by the
Optopus plates. It is clear from this Figure that the coverage of the
cluster galaxy distributions by several relatively small
($33\arcmin$) aperture plates introduces a special selection filter.
For each cluster the plates were positioned to optimize both the
number of measured redshifts (where possible taking into account the
availability of redshifts in the literature) as well as the area
covered. In most cases the positioning of the plates simply
reflects the distribution of galaxies in the cluster. The selection
functions displayed in Figure~10 are, to some extent, arbitrary and
reflect our choice to try and measure as many bright galaxies in a
large, not necessarily circular area, rather than going to fainter
magnitudes in a smaller, circular area. For several applications the
spatial filter does not affect the analysis. If it does play a r\^ole,
e.g.\ when the total luminosity inside a projected distance from the
centre or the uniformity of
the magnitude limit becomes important, the spatial filters are known
in sufficient detail that they can be taken into account.

\section{Discussion and Conclusions}

Even without a detailed analysis of the properties of the systems
that we identified, a few conclusions can already be drawn from the
information in Table~6 and the data displayed in Figure~7.

First, it appears that the large majority of the $R\geq 1$ ACO
clusters with $z\leq 0.1$ (or with $m_{10} > 16.9$) that we observed
indeed correspond to physical systems that are compact in redshift.
Second, the effects of superposition are not negligible, but should
at the same time not be exaggerated. Even for the clusters with only
a relatively small number of redshifts available, most of the
redshifts are generally contained within the dominant system. In
other words: it does not happen very often that an apparently rich
cluster in the ACO catalogue turns out to be the result of a
superpositon of two, about equally rich systems. Only A151 (for which
this was already known) and A2819 are good examples of this kind of
superposition among the clusters with a large number of measured
redshifts. In the class of the clusters with a more modest number of
available redshifts A2426, A2500, A2778, A2871, A3108 and A3703 may
be cases of clusters of which the apparent richness has probably been
boosted significantly as a result of the superposition of two about
equally rich systems. In quite a few of these cases, the velocity
difference between the two systems is not larger than 3000 to 4000
km/s.

The summary of our programme in Figures~7, 9 and 10 immediately
suggests several questions that one can ask from these data, either
in isolation or in combination with data in the literature; and we
are presently pursueing several of these. First, and perhaps most
obviously, the properties of the dominant groups will be studied.
This involves the determination and analysis of the distribution of
velocity dispersions for a complete, volume-limited sample of rich
clusters (Mazure et al.\ 1995), as well as the detailed analysis of
the structure of the phase-space of the individual clusters. Another
aspect of our dataset that is immediately apparent from Figure~7
concerns the information it contains on the large-scale structure in
the Universe. Although it has become customary to discuss the
redshift distributions in very deep, randomly positioned pencil beams
for that purpose (see e.g.\ Broadhurst et al.\ 1990), our 107 pencil
beams in the directions of rich clusters can give complementary
information on the characteristics of large-scale structure. Finally,
the comparison between the dynamics of galaxies with and without
emission lines that could be detected in our observations is
interesting and in progress.

The above list of possible uses that our data can be put to is not
meant to be complete. Clearly, there is interest in comparing the
galaxy kinematics with information from X-ray imaging and
spectroscopy. It will also be very interesting to compare our data to
the predictions of sufficiently realistic N-body simulations of
clusters, in particular to those in which the modeling of the
formation and evolution of galaxies is attempted (e.g.\ by Van
Kampen, 1994 and 1995).

We are still working on several questions for which the present
dataset provides unique new information. The entire dataset will
however be made public in the near future.

In this paper we have concentrated on the properties of our dataset.
We have described the methods that we used to generate it, and we
have discussed its reliability which, at an overall value of 0.98 for
the 5634 redshifts in the catalogue, is quite satisfactory. Finally,
we have used our data to study the effects of field contamination and
superposition, for our sample of nearby and rich Abell clusters. We
conclude that, for such a subset of the ACO catalogue, these effects
are probably sufficiently small that they do not preclude its use for
meaningful statistical studies.

\begin{table}
\caption{Properties of the 220 systems with N$\geq$4 found in the 107 lines
         of sight towards the ACO clusters in the ENACS.}
\begin{flushleft}
\begin{tabular}{lrcccrrl}
\hline
\noalign{\smallskip}
 ACO & N$_{\rm z}$ & z$_{\rm min}$ & z$_{\rm max}$ & $\avrg{\rm z}$ &
 \multicolumn{3}{c}{Literature} \\
\noalign{\smallskip}
& & & & & N$_{\rm z}$ & $\avrg{\rm z}$ & ref. \\
\noalign{\smallskip}
\hline
\noalign{\smallskip}
 A0013 &   4\ \ \ \ & 0.026 & 0.027 & 0.027 &      &       &     \\
       &  37\ \ \ \ & 0.089 & 0.100 & 0.094 &      &       &     \\
 A0087 &  27\ \ \ \ & 0.049 & 0.060 & 0.055 &      &       &     \\
       &   8\ \ \ \ & 0.076 & 0.079 & 0.077 &      &       &     \\
 A0118 &   4\ \ \ \ & 0.058 & 0.059 & 0.059 &      &       &     \\
       &  30\ \ \ \ & 0.110 & 0.120 & 0.115 &      &       &     \\
 A0119 & 104\ \ \ \ & 0.037 & 0.053 & 0.044 &   73 & 0.045 & 1   \\
       &   4\ \ \ \ & 0.139 & 0.140 & 0.140 &      &       &     \\
 A0151 &  25$^{m1}$ & 0.039 & 0.044 & 0.041 &    8 & 0.041 & 2   \\
       &  46$^{m2}$ & 0.048 & 0.058 & 0.053 &   37 & 0.054 & 2   \\
       &  35\ \ \ \ & 0.095 & 0.107 & 0.100 &   10 & 0.110 & 2   \\
 A0168 &   4\ \ \ \ & 0.017 & 0.019 & 0.018 &   22 & 0.018 & 3   \\
       &  76\ \ \ \ & 0.042 & 0.049 & 0.045 &   28 & 0.045 & 3   \\
       &   4\ \ \ \ & 0.069 & 0.074 & 0.072 &      &       &     \\
       &   7\ \ \ \ & 0.089 & 0.092 & 0.090 &      &       &     \\
 A0229 &  32\ \ \ \ & 0.107 & 0.119 & 0.113 &      &       &     \\
 A0295 &  30\ \ \ \ & 0.041 & 0.045 & 0.043 &   28 & 0.042 & 3   \\
       &   5\ \ \ \ & 0.100 & 0.102 & 0.102 &      &       &     \\
 A0303 &   4\ \ \ \ & 0.058 & 0.061 & 0.059 &      &       &     \\
 A0367 &  27$^{b3}$ & 0.084 & 0.098 & 0.091 &      & 0.088 & 4   \\
 A0380 &   4\ \ \ \ & 0.101 & 0.103 & 0.102 &      &       &     \\
       &  25\ \ \ \ & 0.130 & 0.141 & 0.134 &      & 0.135 & 4   \\
 A0420 &  19\ \ \ \ & 0.079 & 0.088 & 0.086 &      &       &     \\
       &   6\ \ \ \ & 0.118 & 0.120 & 0.119 &      &       &     \\
 A0514 &  90\ \ \ \ & 0.066 & 0.080 & 0.072 &    2 & 0.073 & 5   \\
       &   4$^*\;$\ \ & 0.084 & 0.089 & 0.085 &      &       &     \\
       &   8\ \ \ \ & 0.107 & 0.112 & 0.110 &      &       &     \\
 A0524 &  10\ \ \ \ & 0.055 & 0.062 & 0.056 &      &       &     \\
       &  26\ \ \ \ & 0.073 & 0.081 & 0.078 &      &       &     \\
 A0543 &  10\ \ \ \ & 0.082 & 0.088 & 0.085 &      &       &     \\
       &   9\ \ \ \ & 0.166 & 0.176 & 0.171 &      &       &     \\
 A0548 &   4$^*\;$\ \ & 0.030 & 0.032 & 0.031 &      &       &     \\
       & 237\ \ \ \ & 0.035 & 0.051 & 0.042 &  133 & 0.041 & 5   \\
       &   9\ \ \ \ & 0.057 & 0.067 & 0.063 &      &       &     \\
       &  14$^{m1}$ & 0.080 & 0.094 & 0.087 &      &       &     \\
       &  21$^{m2}$ & 0.098 & 0.104 & 0.101 &      &       &     \\
       &   4\ \ \ \ & 0.135 & 0.139 & 0.138 &      &       &     \\
 A0754 &  39\ \ \ \ & 0.044 & 0.061 & 0.055 &   86 & 0.053 & 5   \\
 A0957 &  34\ \ \ \ & 0.038 & 0.051 & 0.045 &   36 & 0.045 & 3   \\
       &            &       &       &       &   23 & 0.044 & 6   \\
 A0978 &  63\ \ \ \ & 0.044 & 0.059 & 0.054 &    2 & 0.053 & 5   \\
 A1069 &  35\ \ \ \ & 0.053 & 0.070 & 0.065 &    2 & 0.063 & 5   \\
       &   4$^*\;$\ \ & 0.113 & 0.115 & 0.114 &      &       &     \\
 A1809 &  30$^{b2}$ & 0.073 & 0.084 & 0.080 &   52 & 0.079 & 7   \\
 A2040 &  37\ \ \ \ & 0.041 & 0.050 & 0.046 &   10 & 0.034 & 3   \\
       &            &       &       &       &   21 & 0.046 & 3   \\
 A2048 &   7\ \ \ \ & 0.037 & 0.042 & 0.040 &      &       &     \\
       &  25\ \ \ \ & 0.094 & 0.103 & 0.097 &    1 & 0.095 & 5   \\
 A2052 &  35\ \ \ \ & 0.030 & 0.045 & 0.035 &   61 & 0.035 & 9   \\
 A2353 &  24\ \ \ \ & 0.117 & 0.125 & 0.121 &      &       &     \\
 A2354 &   4\ \ \ \ & 0.038 & 0.039 & 0.038 &      &       &     \\
       &   5\ \ \ \ & 0.088 & 0.093 & 0.090 &      &       &     \\
\noalign{\smallskip}
\hline
\end{tabular}
\end{flushleft}
\end{table}

\begin{table}
\addtocounter{table}{-1}
\caption[]{$-$ continued $-$}
\begin{flushleft}
\begin{tabular}{lrcccrrl}
\hline
\noalign{\smallskip}
 ACO & N$_{\rm z}$ & z$_{\rm min}$ & z$_{\rm max}$ & $\avrg{\rm z}$ &
 \multicolumn{3}{c}{Literature} \\
\noalign{\smallskip}
& & & & & N$_{\rm z}$ & $\avrg{\rm z}$ & ref. \\
\noalign{\smallskip}
\hline
\noalign{\smallskip}
 A2361 &  24\ \ \ \ & 0.059 & 0.063 & 0.061 &    2 & 0.061 & 5   \\
       &   4$^*\;$\ \ & 0.200 & 0.203 & 0.201 &      &       &     \\
 A2362 &  17\ \ \ \ & 0.060 & 0.064 & 0.061 &    2 & 0.061 & 5   \\
       &   4\ \ \ \ & 0.131 & 0.134 & 0.133 &      &       &     \\
 A2383 &   5\ \ \ \ & 0.057 & 0.061 & 0.058 &      &       &     \\
       &   4\ \ \ \ & 0.113 & 0.118 & 0.116 &      &       &     \\
       &   6\ \ \ \ & 0.129 & 0.132 & 0.130 &      &       &     \\
 A2401 &  23\ \ \ \ & 0.054 & 0.061 & 0.057 &      &       &     \\
       &   5\ \ \ \ & 0.092 & 0.095 & 0.093 &      &       &     \\
 A2426 &  11$^{m1}$ & 0.086 & 0.089 & 0.088 &      &       &     \\
       &  15$^{m2}$ & 0.093 & 0.102 & 0.098 &      &       &     \\
 A2436 &   4\ \ \ \ & 0.056 & 0.061 & 0.058 &      &       &     \\
       &  14\ \ \ \ & 0.087 & 0.095 & 0.091 &      &       &     \\
 A2480 &  11\ \ \ \ & 0.067 & 0.077 & 0.072 &      & 0.071 & 4   \\
 A2500 &  12\ \ \ \ & 0.077 & 0.080 & 0.078 &      &       &     \\
       &  13\ \ \ \ & 0.087 & 0.092 & 0.090 &      &       &     \\
       &   4\ \ \ \ & 0.172 & 0.174 & 0.173 &      &       &     \\
 A2502 &            &       &       &       &      &       &     \\
 A2569 &  36\ \ \ \ & 0.077 & 0.085 & 0.081 &      &       &     \\
 A2644 &   4\ \ \ \ & 0.060 & 0.061 & 0.060 &      &       &     \\
       &  12\ \ \ \ & 0.068 & 0.070 & 0.069 &      &       &     \\
       &   7\ \ \ \ & 0.133 & 0.135 & 0.134 &      &       &     \\
 A2715 &   7\ \ \ \ & 0.054 & 0.060 & 0.055 &      &       &     \\
       &   7\ \ \ \ & 0.096 & 0.098 & 0.098 &      &       &     \\
       &  14\ \ \ \ & 0.111 & 0.120 & 0.114 &      &       &     \\
 A2717 &  40\ \ \ \ & 0.042 & 0.052 & 0.049 &   33 & 0.049 & 8   \\
       &   5\ \ \ \ & 0.072 & 0.073 & 0.073 &      &       &     \\
 A2734 &   5\ \ \ \ & 0.026 & 0.027 & 0.026 &      &       &     \\
       &  83\ \ \ \ & 0.054 & 0.067 & 0.062 &      & 0.063 & 4   \\
       &   4\ \ \ \ & 0.119 & 0.120 & 0.119 &      &       &     \\
       &   6$^*\;$\ \ & 0.137 & 0.141 & 0.141 &      &       &     \\
 A2755 &  22\ \ \ \ & 0.091 & 0.101 & 0.095 &      & 0.095 & 4   \\
       &  10\ \ \ \ & 0.117 & 0.124 & 0.121 &      &       &     \\
 A2764 &  19\ \ \ \ & 0.066 & 0.076 & 0.071 &      & 0.064 & 10  \\
 A2765 &  16$^{m1}$ & 0.076 & 0.083 & 0.080 &      &       &     \\
       &   4$^{m2}$ & 0.088 & 0.093 & 0.090 &      &       &     \\
 A2778 &   7\ \ \ \ & 0.075 & 0.079 & 0.077 &      &       &     \\
       &  17$^{m1}$ & 0.097 & 0.108 & 0.102 &      & 0.103 & 4   \\
       &  10$^{m2}$ & 0.115 & 0.121 & 0.119 &      & 0.119 & 4   \\
 A2799 &  36\ \ \ \ & 0.060 & 0.067 & 0.063 &      & 0.062 & 4   \\
 A2800 &  34\ \ \ \ & 0.059 & 0.067 & 0.064 &      &       &     \\
 A2819 &  49$^{m1}$ & 0.071 & 0.078 & 0.075 &      &       &     \\
       &  43$^{m2}$ & 0.082 & 0.090 & 0.087 &      & 0.087 & 4   \\
       &   4\ \ \ \ & 0.106 & 0.108 & 0.106 &      &       &     \\
       &   4\ \ \ \ & 0.131 & 0.137 & 0.133 &      &       &     \\
       &  13\ \ \ \ & 0.157 & 0.163 & 0.160 &      &       &     \\
 A2854 &  22\ \ \ \ & 0.060 & 0.064 & 0.061 &      &       &     \\
 A2871 &  14$^{m1}$ & 0.112 & 0.116 & 0.114 &      &       &     \\
       &  18$^{m2}$ & 0.120 & 0.130 & 0.122 &      &       &     \\
 A2911 &   7\ \ \ \ & 0.020 & 0.022 & 0.020 &      &       &     \\
       &  31\ \ \ \ & 0.075 & 0.086 & 0.081 &      & 0.079 & 4   \\
       &   4\ \ \ \ & 0.130 & 0.133 & 0.131 &      &       &     \\
 A2915 &   4$^*\;$\ \ & 0.086 & 0.087 & 0.086 &      &       &     \\
\noalign{\smallskip}
\hline
\end{tabular}
\end{flushleft}
\end{table}

\begin{table}
\addtocounter{table}{-1}
\caption[]{$-$ continued $-$}
\begin{flushleft}
\begin{tabular}{lrcccrrl}
\hline
\noalign{\smallskip}
 ACO & N$_{\rm z}$ & z$_{\rm min}$ & z$_{\rm max}$ & $\avrg{\rm z}$ &
 \multicolumn{3}{c}{Literature} \\
\noalign{\smallskip}
& & & & & N$_{\rm z}$ & $\avrg{\rm z}$ & ref. \\
\noalign{\smallskip}
\hline
\noalign{\smallskip}
 A2923 &   5\ \ \ \ & 0.016 & 0.019 & 0.017 &      &       &     \\
       &  16\ \ \ \ & 0.071 & 0.074 & 0.071 &      &       &     \\
       &   4\ \ \ \ & 0.127 & 0.130 & 0.128 &      &       &     \\
 A2933 &   9\ \ \ \ & 0.090 & 0.094 & 0.093 &      &       &     \\
 A2954 &   6\ \ \ \ & 0.056 & 0.057 & 0.057 &      &       &     \\
       &   5\ \ \ \ & 0.125 & 0.127 & 0.126 &      &       &     \\
 A3009 &  12\ \ \ \ & 0.063 & 0.070 & 0.065 &      & 0.075 & 4   \\
 A3093 &   5\ \ \ \ & 0.063 & 0.067 & 0.064 &      & 0.064 & 4   \\
       &  22\ \ \ \ & 0.080 & 0.086 & 0.083 &      &       &     \\
       &   6\ \ \ \ & 0.113 & 0.118 & 0.115 &      &       &     \\
 A3094 &  69\ \ \ \ & 0.063 & 0.075 & 0.068 &      &       &     \\
       &   5\ \ \ \ & 0.106 & 0.108 & 0.108 &      &       &     \\
       &  12\ \ \ \ & 0.136 & 0.143 & 0.139 &      &       &     \\
 A3108 &   7\ \ \ \ & 0.060 & 0.063 & 0.063 &      &       &     \\
       &   5$^*\;$\ \ & 0.081 & 0.082 & 0.082 &      &       &     \\
 A3111 &  35\ \ \ \ & 0.074 & 0.084 & 0.078 &      & 0.080 & 4   \\
 A3112 &  77\ \ \ \ & 0.059 & 0.082 & 0.075 &      & 0.076 & 4   \\
       &   4\ \ \ \ & 0.090 & 0.092 & 0.090 &      &       &     \\
       &  14\ \ \ \ & 0.128 & 0.140 & 0.132 &      &       &     \\
 A3122 &  92\ \ \ \ & 0.055 & 0.070 & 0.064 &      &       &     \\
       &   8\ \ \ \ & 0.114 & 0.120 & 0.117 &      &       &     \\
       &  10\ \ \ \ & 0.148 & 0.152 & 0.150 &      &       &     \\
 A3128 &  12$^*\;$\ \ & 0.039 & 0.047 & 0.039 &      &       &     \\
       & 158\ \ \ \ & 0.051 & 0.071 & 0.060 &   43 & 0.059 & 8   \\
       &  11$^*\;$\ \ & 0.075 & 0.078 & 0.077 &      &       &     \\
       &   4\ \ \ \ & 0.106 & 0.108 & 0.107 &      &       &     \\
 A3141 &  15\ \ \ \ & 0.101 & 0.109 & 0.105 &      & 0.107 & 4   \\
 A3142 &12$^{b2}$ & 0.062 & 0.069 & 0.066 &      &       &     \\
       &  21\ \ \ \ & 0.096 & 0.107 & 0.103 &      & 0.103 & 4   \\
 A3144 &            &       &       &       &   16 & 0.045 & 4   \\
 A3151 &  38\ \ \ \ & 0.059 & 0.072 & 0.068 &      & 0.068 & 4   \\
 A3158 & 105\ \ \ \ & 0.052 & 0.067 & 0.059 &      & 0.058 & 10  \\
       &   4$^*\;$\ \ & 0.072 & 0.074 & 0.074 &      &       &     \\
       &   4$^*\;$\ \ & 0.101 & 0.102 & 0.102 &      &       &     \\
 A3194 &  32\ \ \ \ & 0.093 & 0.102 & 0.097 &      & 0.098 & 4   \\
 A3202 &  27\ \ \ \ & 0.066 & 0.072 & 0.069 &      &       &     \\
 A3223 &  81\ \ \ \ & 0.054 & 0.075 & 0.060 &      & 0.063 & 4   \\
       &   8\ \ \ \ & 0.109 & 0.111 & 0.110 &      &       &     \\
       &   8\ \ \ \ & 0.135 & 0.140 & 0.137 &      &       &     \\
 A3264 &   5\ \ \ \ & 0.095 & 0.099 & 0.098 &      &       &     \\
 A3301 &   5\ \ \ \ & 0.053 & 0.057 & 0.054 &      &       &     \\
 A3341 &  64\ \ \ \ & 0.034 & 0.043 & 0.038 &      & 0.037 & 4   \\
       &  15\ \ \ \ & 0.075 & 0.082 & 0.078 &      &       &     \\
       &  18\ \ \ \ & 0.112 & 0.117 & 0.115 &      &       &     \\
       &   7\ \ \ \ & 0.130 & 0.134 & 0.131 &      &       &     \\
       &   5\ \ \ \ & 0.151 & 0.154 & 0.154 &      &       &     \\
 A3354 &   4\ \ \ \ & 0.035 & 0.037 & 0.036 &      &       &     \\
       & 5$^{m1}$   & 0.042 & 0.044 & 0.044 &      &       &     \\
       &58$^{m2}$   & 0.055 & 0.062 & 0.059 &      &       &     \\
       &   4\ \ \ \ & 0.079 & 0.083 & 0.082 &      &       &     \\
       &   6\ \ \ \ & 0.115 & 0.119 & 0.118 &      &       &     \\
       &   5\ \ \ \ & 0.132 & 0.135 & 0.135 &      &       &     \\
       &   5\ \ \ \ & 0.144 & 0.146 & 0.145 &      &       &     \\
       &   5\ \ \ \ & 0.160 & 0.167 & 0.164 &      &       &     \\
       &   7\ \ \ \ & 0.194 & 0.197 & 0.195 &      &       &     \\
\noalign{\smallskip}
\hline
\end{tabular}
\end{flushleft}
\end{table}

\begin{table}
\addtocounter{table}{-1}
\caption[]{$-$ continued $-$}
\begin{flushleft}
\begin{tabular}{lrcccrrl}
\hline
\noalign{\smallskip}
 ACO & N$_{\rm z}$ & z$_{\rm min}$ & z$_{\rm max}$ & $\avrg{\rm z}$ &
 \multicolumn{3}{c}{Literature} \\
\noalign{\smallskip}
& & & & & N$_{\rm z}$ & $\avrg{\rm z}$ & ref. \\
\noalign{\smallskip}
\hline
\noalign{\smallskip}
 A3365 &  32\ \ \ \ & 0.088 & 0.101 & 0.092 &      &       &     \\
 A3528 &  28\ \ \ \ & 0.047 & 0.058 & 0.054 &      &       &     \\
       &   9\ \ \ \ & 0.069 & 0.077 & 0.073 &      &       &     \\
 A3558 &   4\ \ \ \ & 0.030 & 0.033 & 0.032 &      &       &     \\
       &  75\ \ \ \ & 0.040 & 0.056 & 0.048 &  267 & 0.048 & 11  \\
 A3559 &   7\ \ \ \ & 0.013 & 0.015 & 0.014 &      &       &     \\
       &  39\ \ \ \ & 0.043 & 0.050 & 0.047 &      &       &     \\
       &   6\ \ \ \ & 0.072 & 0.078 & 0.077 &      &       &     \\
       &  11\ \ \ \ & 0.110 & 0.116 & 0.113 &      &       &     \\
 A3562 & 118\ \ \ \ & 0.035 & 0.055 & 0.048 &      &       &     \\
 A3651 &  79\ \ \ \ & 0.053 & 0.065 & 0.060 &      &       &     \\
       &   5\ \ \ \ & 0.100 & 0.101 & 0.101 &      &       &     \\
 A3667 & 103\ \ \ \ & 0.046 & 0.065 & 0.056 &  122 & 0.055 & 12  \\
       &   5\ \ \ \ & 0.097 & 0.103 & 0.099 &      &       &     \\
 A3677 &   8\ \ \ \ & 0.086 & 0.096 & 0.091 &      &       &     \\
 A3682 &10$^{b2}$ & 0.086 & 0.093 & 0.092 &      &       &     \\
 A3691 &  33\ \ \ \ & 0.081 & 0.094 & 0.087 &      &       &     \\
 A3693 &  16\ \ \ \ & 0.088 & 0.094 & 0.091 &      &       &     \\
       &   9\ \ \ \ & 0.120 & 0.129 & 0.124 &      &       &     \\
 A3695 &  81\ \ \ \ & 0.082 & 0.099 & 0.089 &      &       &     \\
       &   7\ \ \ \ & 0.130 & 0.132 & 0.131 &      &       &     \\
 A3696 &12$^{b2}$ & 0.086 & 0.092 & 0.088 &      &       &     \\
 A3703 &  18\ \ \ \ & 0.070 & 0.075 & 0.074 &      & 0.071 & 10  \\
       &  13\ \ \ \ & 0.089 & 0.098 & 0.091 &      &       &     \\
 A3705 &   4\ \ \ \ & 0.047 & 0.048 & 0.047 &      &       &     \\
       &  29\ \ \ \ & 0.084 & 0.097 & 0.090 &   40 & 0.090 & 8   \\
 A3733 &  41\ \ \ \ & 0.033 & 0.043 & 0.039 &   17 & 0.046 & 4   \\
 A3744 &  71\ \ \ \ & 0.034 & 0.049 & 0.038 &      &       &     \\
       &   5\ \ \ \ & 0.062 & 0.067 & 0.065 &      &       &     \\
 A3764 &  38\ \ \ \ & 0.072 & 0.084 & 0.076 &      &       &     \\
 A3781 &   4\ \ \ \ & 0.057 & 0.058 & 0.057 &      &       &     \\
       &   4\ \ \ \ & 0.071 & 0.074 & 0.073 &      &       &     \\
 A3795 &  13\ \ \ \ & 0.086 & 0.091 & 0.089 &      &       &     \\
 A3799 &  10\ \ \ \ & 0.043 & 0.047 & 0.045 &      &       &     \\
 A3806 &   9\ \ \ \ & 0.053 & 0.055 & 0.054 &      &       &     \\
       &  99\ \ \ \ & 0.065 & 0.086 & 0.076 &      &       &     \\
       &   4\ \ \ \ & 0.137 & 0.139 & 0.138 &      &       &     \\
 A3809 &  94\ \ \ \ & 0.057 & 0.072 & 0.062 &      & 0.062 & 4   \\
       &   4\ \ \ \ & 0.090 & 0.091 & 0.091 &      &       &     \\
       &  10\ \ \ \ & 0.108 & 0.112 & 0.110 &      &       &     \\
       &  11\ \ \ \ & 0.139 & 0.148 & 0.141 &      &       &     \\
       &   4$^*\;$\ \ & 0.152 & 0.156 & 0.152 &      &       &     \\
 A3822 &   4$^*\;$\ \ & 0.037 & 0.040 & 0.039 &      &       &     \\
       &   4\ \ \ \ & 0.052 & 0.053 & 0.052 &      &       &     \\
       &  84\ \ \ \ & 0.066 & 0.083 & 0.076 &      &       &     \\
       &   4\ \ \ \ & 0.099 & 0.105 & 0.102 &      &       &     \\
 A3825 &  61\ \ \ \ & 0.067 & 0.080 & 0.075 &      &       &     \\
       &17$^{m1}$ & 0.097 & 0.112 & 0.104 &      &       &     \\
       & 4$^{m2}$ & 0.116 & 0.121 & 0.119 &      &       &     \\
 A3827 &  20\ \ \ \ & 0.093 & 0.108 & 0.098 &      & 0.099 & 10  \\
 A3864 &   6\ \ \ \ & 0.075 & 0.079 & 0.077 &      &       &     \\
       &  32\ \ \ \ & 0.095 & 0.109 & 0.102 &      &       &     \\
\noalign{\smallskip}
\hline
\end{tabular}
\end{flushleft}
\end{table}

\begin{table}
\addtocounter{table}{-1}
\caption[]{$-$ continued $-$}
\begin{flushleft}
\begin{tabular}{lrcccrrl}
\hline
\noalign{\smallskip}
 ACO & N$_{\rm z}$ & z$_{\rm min}$ & z$_{\rm max}$ & $\avrg{\rm z}$ &
 \multicolumn{3}{c}{Literature} \\
\noalign{\smallskip}
& & & & & N$_{\rm z}$ & $\avrg{\rm z}$ & ref. \\
\noalign{\smallskip}
\hline
\noalign{\smallskip}
 A3879 &   5\ \ \ \ & 0.050 & 0.051 & 0.050 &      &       &     \\
       &  45\ \ \ \ & 0.059 & 0.074 & 0.067 &      & 0.068 &  4  \\
       &   7\ \ \ \ & 0.095 & 0.099 & 0.097 &      &       &     \\
       &   5\ \ \ \ & 0.127 & 0.131 & 0.130 &      &       &     \\
 A3897 &  10\ \ \ \ & 0.071 & 0.077 & 0.073 &      &       &     \\
 A3921 &  32\ \ \ \ & 0.086 & 0.101 & 0.094 &      & 0.096 & 4   \\
       &   4\ \ \ \ & 0.133 & 0.136 & 0.134 &      &       &     \\
 A4008 &  27\ \ \ \ & 0.052 & 0.057 & 0.055 &      &       &     \\
       &   7\ \ \ \ & 0.102 & 0.114 & 0.107 &      &       &     \\
 A4010 &  30\ \ \ \ & 0.091 & 0.100 & 0.096 &      &       &     \\
 A4053 &   9\ \ \ \ & 0.049 & 0.052 & 0.050 &      &       &     \\
       &  17\ \ \ \ & 0.066 & 0.075 & 0.072 &      &       &     \\
\noalign{\smallskip}
\hline
\end{tabular}
\end{flushleft}
{{\bf Notes:}
Systems adjacent in redshift that are indicated by $m1$, $m2$, $m3$ etc., are
merged into one group by the method of ZHG. Systems marked by $b2$ or $b3$
are broken up in two or three groups respectively by the ZHG method.
An $*$ indicates that the system is not found by the ZHG method.
}\\
{{\bf References:}
 1. Fabricant et al.\ (1993); 2. Proust et al.\ (1992);
 3. Zabludoff et al.\ (1990); 4. Dalton et (1994); 5. Struble \& Rood (1991);
 6. Capelato et al.\ (1991); 7. Hill \& Oegerle (1993);
 8. Colless \& Hewett (1987); 9. Malumuth et al.\ (1992);
10. Andernach (priv. comm.) 11. Bardelli et al.\ (1994);
12. Sodr\'e et al.\ (1992)}
\end{table}

\begin{acknowledgements}
{We gratefully acknowledge the support given to this project by ESO,
first by accepting it as a Key-programme, and secondly by allocating
the large amount of telescope time without which this project would
not have been possible. We extend special thanks to G.\ Avila, who
has been very helpful in many aspects of the observational program,
such as the production of the aperture plates, and the optimization
of the performance of the Optopus multi-fibre system. At the 3.6-m
telescope we were very ably assisted by G.\ Roman, A.\ Alvarez,
M.\ Bahamondes, and L.\ Ramirez. PK acknowledges very useful
discussions with R.S.\ Le\ Poole on several aspects of the data
analysis. The cooperation between the members of the project was
financially supported by the following organizations: INSU, GR
Cosmologie, Univ. de Provence, Univ. de Montpellier (France),
CNRS-NWO (France and the Netherlands), Leiden Observatory, Leids
Kerkhoven-Bosscha Fonds (the Netherlands), Univ. of Trieste,
Univ.  of Bologna (Italy), the Swiss National Science Foundation, the
Ministerio de Educacion y Ciencia (Spain), CNRS-CSIC (France and
Spain) and by the EC HCM programme. Finally, we acknowledge several
useful comments of the referee, L.\ Guzzo.}
\end{acknowledgements}
\vfill
\eject

\appendix
\section{The Estimation of the Reliability of our Redshifts}

Here we present some details of the evidence on which we based our
estimates of the reliability of the redshifts. All the evidence is
empirical and based on independent multiple measurements. First, we
use two sets of consecutive Optopus exposures (with only the CCD
read-out in between), which were reduced separately, to estimate the
S/N-ratio in the correlation function above which the reliability is
essentially 1.0. Then we use all available pairs of multiple
measurements of {\em accepted} absorption-line redshift estimates to
deduce the actual reliability of the redshifts in our catalogue. A
similar analysis is made for the multiple measurements of
emission-line redshifts. Finally, we use all galaxies for which both
an (independently measured) absorption- and emission-line redshift
were obtained to refine our estimates of the reliability of the
different types of redshift estimate.

The two sets of consecutive double exposures, which were reduced
completely separately (contrary to the normal procedure for double
exposures, which were combined before calibration and reduction), with
46 and 47 spectra respectively, yielded 93 pairs of redshifts. Of
these, 23 were found to be discordant, and in all discordant pairs at
least one of the redshifts had a S/N-ratio of the correlation peak
less than 3.0. The statistics in Table~7 shows that for a S/N-ratio
$\ge$ 3.0 the reliability of a redshift estimate is essentially 1.0,
while for S/N-ratios $<$ 3.0 the average reliability is of the order
of 0.60$\pm$ 0.05. Note that for this analysis {\em all} redshift
pairs were used (i.e.\ the plausibility of individual redshift
estimates was not judged), as we are interested here in the influence
of the noise in the correlation function. From Table~7 we conclude
that the S/N-ratio below which the reliability of a redshift estimate
starts to decline (fairly rapidly) is about 3.0. In the following we
will therefore use this value to separate redshifts with identical
high reliability (equal to 1.0) from redshifts with lower
reliabilities (between 0.0 and 1.0).

\begin{table}
\caption[]{The reliability of the redshifts as inferred from blindly accepted
estimates in 2 consecutive exposures.}
\begin{flushleft}
\begin{tabular}{rrrrr}
\hline\hline
\noalign{\smallskip}
$S/N_1$ & $S/N_2$  & \# pairs & concordant & discordant \\
\noalign{\smallskip}
\hline
\noalign{\smallskip}
 $\ge$3.0 & $\ge$3.0 & 52 & 52 &  0 \\
 $\ge$3.0 &   $<$3.0 & 10 &  7 &  3 \\
   $<$3.0 &   $<$3.0 & 31 & 11 & 20 \\
& &\hrulefill&\hrulefill&\hrulefill\\
\multicolumn{2}{l}{total pairs:} & 93 & 70 & 23 \\
\noalign{\smallskip}
\hline\hline
\end{tabular}
\end{flushleft}
\end{table}

Rather than accept all redshift estimates blindly (as was done for
the statistics in Table~7), each and every redshift estimate was
checked for plausibility by simultaneous visual inspection of the
spectrum and the correlation function. The reliability of essentially 1.0
for the S/N $\ge$ 3.0 redshifts was confirmed by the visual
inspection and none of these redshifts was rejected. However, quite a
few of the S/N $<$ 3.0 redshifts were judged very improbable and
rejected. Therefore, the reliability of the {\em accepted} redshifts with
S/N $<$ 3.0 is very likely to be much higher than the value of 0.6
implied by the data in Table~7.

\begin{table}
\caption[]{The reliability of the redshifts as inferred from pairs of
estimates accepted after inspection of correlation function and spectrum}
\begin{flushleft}
\begin{tabular}{rrrrr}
\hline\hline
\noalign{\smallskip}
$S/N_1$ & $S/N_2$  & \# pairs & concordant & discordant \\
\noalign{\smallskip}
\hline
\noalign{\smallskip}
 $\ge$3.0 & $\ge$3.0 & 265 & 265 &  0 \\
 $\ge$3.0 &   $<$3.0 &  84 &  80 &  4 \\
   $<$3.0 &   $<$3.0 &  43 &  39 &  4 \\
& &\hrulefill&\hrulefill&\hrulefill\\
\multicolumn{2}{l}{total pairs:} & 392 & 384 & 8 \\
\noalign{\smallskip}
\hline\hline
\end{tabular}
\end{flushleft}
\end{table}

We have used the multiple measurements of redshifts to estimate the
actual reliability of the {\em accepted} S/N $<$ 3.0 redshifts. For 360
galaxies at least two independent redshift estimates were accepted;
in 22 cases three, and in 5 cases even four redshifts were available
for a given galaxy. Thus, we have 392 independent pairs for which we
can carry out the same analysis as for the 97 pairs in the two double
exposures discussed above. In Table~8 we show the statistics for the
392 pairs. From these we conclude that the reliability of the
S/N-ratio $\ge$ 3.0 redshifts is indeed 1.00. The average reliability of
the {\em accepted} S/N-ratio $<$ 3.0 redshifts turns out to be
considerably higher than 0.6, namely about 0.95. This illustrates the
great value of the visual inspection, which seems to have reduced the
fraction of erroneous redshifts by almost 90 per cent.

\begin{table}
\caption[]{The reliability of redshift estimates based on emission lines.}
\begin{flushleft}
\begin{tabular}{rrrrr}
\hline\hline
\noalign{\smallskip}
$\#$-lines$_1$ & $\#$-lines$_2$  & \# pairs & concordant & discordant \\
\noalign{\smallskip}
\hline
\noalign{\smallskip}
 multiple & multiple &  30 &  30 &  0 \\
 multiple &   single &  14 &  12 &  2 \\
   single &   single &   3 &   3 &  0 \\
& &\hrulefill&\hrulefill&\hrulefill\\
\multicolumn{2}{l}{total pairs:} & 47 & 45 & 2 \\
\noalign{\smallskip}
\hline\hline
\end{tabular}
\end{flushleft}
\end{table}

The reliability of the emission-line redshifts was estimated in an
analogous way. We have 47 independent pairs of emission-line
redshifts. It seems quite natural to distinguish estimates based on
one line only (whether it be OII, H${\beta}$ or OIII) from estimates
based on a combination of at least two lines. In Table~9 we show the
rather limited statistics. As a first approximation, we conclude that
emission-line redshifts based on at least two lines have a
reliability of essentially 1.00, while those based on a single line
(which does not need to be the OIII doublet) have a reliability
of 0.85 $\pm$ 0.10.

\begin{table}
\caption[]{Agreement between absorption- and emission-line redshifts.}
\begin{flushleft}
\begin{tabular}{llrrr}
\hline\hline
\noalign{\smallskip}
$z_{\rm abs}$ & $z_{\rm emi}$ & \# pairs & concordant & discordant \\
\noalign{\smallskip}
\hline
\noalign{\smallskip}
S/N $\ge$ 3.0 &  $\ge2$ lines & 162 & 159 &   3 \\
              &  1 line       & 162 & 130 &  32 \\
              &               &     &     &     \\
S/N $<$ 3.0   &  $\ge$2 lines & 230 & 210 &  20 \\
              &  1 line       & 112 &  87 &  25 \\
& &\hrulefill&\hrulefill&\hrulefill\\
\multicolumn{2}{l}{total pairs:} & 666 & 586 &  80 \\
\noalign{\smallskip}
\hline\hline
\end{tabular}
\end{flushleft}
\end{table}

Finally, we have analyzed the results for the 666 galaxies with both
an absorption- and an emission-line redshift. In Table~10 we show the
statistics of the concordant and discordant pairs. We interpret the
data in this Table as follows. Accepting the previous result that the
reliability of the absorption-line redshifts with S/N $\ge$ 3.0 is
1.00, we conclude from the upper half of the Table that the
reliability of the `multiple' emission-line redshifts is 0.98.
Similarly, the single emission-line redshifts appear to have a
reliability of about 0.80. Note that both conclusions are completely
consistent with the earlier estimates. The lower half of the Table
then tells one that the reliability of the accepted S/N $<$ 3.0
absorption-line redshifts is 0.95 (again confirming the earlier,
completely independent, estimate). The multiple-line emission-line
redshifts have a reliability of 0.97, and the single-line
emission-line redshifts of 0.81. All these reliability estimates are
believed to be accurate to between 1 and 2 per cent.
\vfill

\begin{thebibliography}{}
\bibitem{} Abell, G.O., 1958, ApJS, 3, 211
\bibitem{} Abell, G.O., Corwin, H.G., Olowin, R.P., 1989, ApJS, 70, 1 (ACO)
\bibitem{} Abt, H.A., Biggs, E.S., 1972, Bibliography of Stellar Radial
Velocities, Latham Process Corp., New York
\bibitem{} Avila, G., D'Odorico, S., Tarenghi, M., Guzzo, L., 1989, ESO
Messenger, No.\ 55, p.\ 62
\bibitem{} Bardelli, S., Zucca, E., Vettolani, et al.\, 1994, MNRAS, 267, 665
\bibitem{} Beers, T.C., Forman, W., Huchra, J.P., Jones, C., Gebhardt, K.,
1991, AJ, 102, 1581
\bibitem{} Biviano, A., Girardi, M., Giuricin, G., Mardirossian, F., Mezzetti,
M., 1992, ApJ, 396, 35
\bibitem{} Briel, U.G., Henry, J.P., 1993, A\&A, 278, 390
\bibitem{} Broadhurst, T.J., Ellis, R.S., Koo, D.C., Szalay, A.S., 1990,
Nature, 343, 726
\bibitem{} Capelato, H.V., Mazure, A., Proust, et al.\, 1991, A\&ASS, 90, 355
\bibitem{} Chincarini, G., Tarenghi, M., Bettis, C., 1981, A\&A, 96, 106
\bibitem{} Colless, M., Hewett, P., 1987, MNRAS, 224, 453
\bibitem{} Colless, M., 1989, MNRAS, 237, 799
\bibitem{} Dalton, G.B., Efstathiou, G., Maddox, S.J., Sutherland, W.J., 1992,
ApJL, 390, L1
\bibitem{} Dalton, G.B., Efstathiou, G., Maddox, S.J., Sutherland, W.J., 1994,
MNRAS, 269, 151
\bibitem{} De Vaucouleurs, G., De Vaucouleurs, A., Corwin, H.G. Jr., et al.\,
1991, Third Reference catalogue of Bright Galaxies, ed. Springer-Verlag
\bibitem{} De Vries, C.P., 1987, PhD thesis, Leiden
\bibitem{} Dressler, A., Shectman, S.A., 1988, AJ, 95, 284
\bibitem{} Faber, S.M., Dressler, A., 1977, AJ, 82, 187
\bibitem{} Fabricant, D., Beers, T.C., Geller, M.J., et al.\, 1986, ApJ, 308,
530
\bibitem{} Fabricant, D., Kurtz, M.J., Geller, M., et al.\, 1993, AJ, 105, 788
\bibitem{} Guzzo, L., Collins, C.A., Nichol, R.C., Lumsden, S.L., 1992, ApJ,
393, L5
\bibitem{} den Hartog, R.H., Katgert, P., 1995, submitted to MNRAS
\bibitem{} Heydon-Dumbleton, N.H., Collins, C.A., MacGillivray, H.T., 1989,
MNRAS, 238, 379
\bibitem{} Hill, J.M., and Oegerle, W.R., 1993, AJ, 106, 831
\bibitem{} Le F\`evre,O., Bijaoui, A., Mathez, G., Picat, J.P., Lelievre, G.,
1986, A\&A, 154, 92
\bibitem{} Lissandrini, C., Cristiani, S., La Franca, F., 1994, PASP, 106, 1157
\bibitem{} Lucey, J.R., 1983, MNRAS, 204, 33
\bibitem{} Lucey, J.R., Dickens, R.J., Mitchell, R.J., Dawe, J.A., 1983, MNRAS,
203, 545
\bibitem{} Lumsden, S.L., Nichol, R.C., Collins, C.A., Guzzo, L., 1992, MNRAS,
258, 1
\bibitem{} Lund, G., 1986, ESO Operating Manual No. 6
\bibitem{} Maddox, S.J., Sutherland, W.J., Efstathiou, G., Loveday, J., 1990,
MNRAS, 243, 692
\bibitem{} Malumuth, E.M., Kriss, G.A., Van Dyke Dixon, W., Ferguson, H.C.,
Ritchie, C., 1992, AJ, 104, 495
\bibitem{} Materne, J., Hopp, U., 1983, A\&A, 124, L13
\bibitem{} Mazure, A., Katgert, P., Den Hartog, R.H., et al.\, 1995, TO BE
FILLED IN BY EDITOR AANDA
\bibitem{} Pierre, M., B\"ohringer, H., Ebeling, H., et al.\, 1994, A\&A, 290,
725
\bibitem{} Pisani, A., 1993, MNRAS, 265, 706
\bibitem{} Proust, D., Quintana, H., Mazure, A., et al.\, 1992, A\&A, 258, 243
\bibitem{} Sarazin, C., 1986, Rev. Mod. Phys., 58,1
\bibitem{} Sodr\'e, L., Capelato, H.V., Steiner, J.E., Proust, D., Mazure, A.,
1992, MNRAS, 259, 233
\bibitem{} Struble, M.F., Rood, H.J., 1991, ApJS, 77, 363
\bibitem{} Sutherland, W., 1988, MNRAS, 234, 159
\bibitem{} Swaans, L., 1981, PhD thesis, Leiden
\bibitem{} Teague, P.F., Carter, D., Gray, P.M., 1990, ApJS, 72, 715
\bibitem{} Tonry, J., Davis, M., 1979, AJ, 84, 1511
\bibitem{} Van Haarlem, M.P., Le Poole, R.S., Katgert, P., Tritton, S., 1991,
MNRAS, 255,
\bibitem{} Van Kampen, E., 1994, PhD thesis, Leiden Observatory
\bibitem{} Van Kampen, E., 1995, MNRAS, 273, 295
\bibitem{} West, M.J., Dekel, A., Oemler, A., 1987, ApJ, 316, 1
\bibitem{} White, S.D.M., 1991, in {\em Large Scale Structures and Peculiar
Motions in the Universe}, eds.\ Latham, D.W. \& Da Costa, L.N., ASP, p285
\bibitem{} White, S.D.M., 1992, in {\em Clusters and Superclusters of
Galaxies}, ed.\ A.C. Fabian, NATO ASI Series C366, Kluwer Acad.\ Publ.\,
Dordrecht
\bibitem{} Zabludoff, A.I., Huchra, J.P., Geller, M.J., 1990, ApJS, 74, 1 (ZHG)
\bibitem{} Zabludoff, A.I., Geller, M.J., Huchra, J.P., Vogeley, M.S., 1993,
AJ, 106, 1301 (ZHGV)
\bibitem{} Zucca, E., Zamorani, G., Scaramella, R., Vettolani, G.,
1993, ApJ, 407, 470
\end{thebibliography}
\end{document}